
\input harvmac

%
\newbox\hdbox%
\newcount\hdrows%
\newcount\multispancount%
\newcount\ncase%
\newcount\ncols
\newcount\nrows%
\newcount\nspan%
\newcount\ntemp%
\newdimen\hdsize%
\newdimen\newhdsize%
\newdimen\parasize%
\newdimen\spreadwidth%
\newdimen\thicksize%
\newdimen\thinsize%
\newdimen\tablewidth%
\newif\ifcentertables%
\newif\ifendsize%
\newif\iffirstrow%
\newif\iftableinfo%
\newtoks\dbt%
\newtoks\hdtks%
\newtoks\savetks%
\newtoks\tableLETtokens%
\newtoks\tabletokens%
\newtoks\widthspec%
%
%
\immediate\write15{%
CP SMSG GJMSINK TEXTABLE --> TABLE MACROS V. 851121 JOB = \jobname%
}%
%
%
\tableinfotrue%
\catcode`\@=11
%
%
\def\tstrut{\vrule height3.1ex depth1.2ex width0pt}%
\def\and{\char`\&}
\def\tablerule{\noalign{\hrule height\thinsize depth0pt}}%
\thicksize=1.5pt
\thinsize=0.6pt
\def\thickrule{\noalign{\hrule height\thicksize depth0pt}}%
\def\ctr#1{\hfil\ #1\hfil}%
%
%
%
%
\tablewidth=-\maxdimen%
\spreadwidth=-\maxdimen%
\def\tabskipglue{0pt plus 1fil minus 1fil}%
%
%
\centertablestrue%
%
%
%
%
\parasize=4in%
\gdef\ARGS{########}
\gdef\headerARGS{####}
\def\@mpersand{&}
{\catcode`\|=13
\gdef\letbarzero{\let|0}
\gdef\letbartab{\def|{&&}}%
\gdef\letvbbar{\let\vb|}%
}
{\catcode`\&=4
\def\ampskip{&\omit\hfil&}
\catcode`\&=13
\let&0
\xdef\letampskip{\def&{\ampskip}}%
\gdef\letnovbamp{\let\novb&\let\tab&}
}
\def\begintable{
   \begingroup%
   \catcode`\|=13\letbartab\letvbbar%
   \catcode`\&=13\letampskip\letnovbamp%
   \def\multispan##1{
      \omit \mscount##1%
      \multiply\mscount\tw@\advance\mscount\m@ne%
      \loop\ifnum\mscount>\@ne \sp@n\repeat%
   }
   \def\|{%
      &\omit\widevline&%
   }%
   \ruledtable
}
\long\def\ruledtable#1\endtable{%
%
%
%
   \offinterlineskip
   \tabskip 0pt
   \def\widevline{\vrule width\thicksize}
   \def\endrow{\@mpersand\omit\hfil\crnorm\@mpersand}%
   \def\crthick{\@mpersand\crnorm\thickrule\@mpersand}%
   \def\crthickneg##1{\@mpersand\crnorm\thickrule
          \noalign{{\skip0=##1\vskip-\skip0}}\@mpersand}%
   \def\crnorule{\@mpersand\crnorm\@mpersand}%
   \def\crnoruleneg##1{\@mpersand\crnorm
          \noalign{{\skip0=##1\vskip-\skip0}}\@mpersand}%
   \let\nr=\crnorule
   \def\endtable{\@mpersand\crnorm\thickrule}%
   \let\crnorm=\cr
%
%
   \edef\cr{\@mpersand\crnorm\tablerule\@mpersand}%
   \def\crneg##1{\@mpersand\crnorm\tablerule
          \noalign{{\skip0=##1\vskip-\skip0}}\@mpersand}%
   \let\ctneg=\crthickneg
   \let\nrneg=\crnoruleneg
   \the\tableLETtokens
%
%
   \tabletokens={&#1}
%
%
   \countROWS\tabletokens\into\nrows%
   \countCOLS\tabletokens\into\ncols%
%
%
   \advance\ncols by -1%
   \divide\ncols by 2%
   \advance\nrows by 1%
%
%
   \iftableinfo %
      \immediate\write16{[Nrows=\the\nrows, Ncols=\the\ncols]}%
   \fi%
%
%
   \ifcentertables
      \ifhmode \par\fi
      \hbox to \hsize{
      \hss
   \else %
      \hbox{%
   \fi
      \vbox{%
         \makePREAMBLE{\the\ncols}
         \edef\next{\preamble}
         \let\preamble=\next
         \makeTABLE{\preamble}{\tabletokens}
      }
      \ifcentertables \hss}\else }\fi
   \endgroup
   \tablewidth=-\maxdimen
   \spreadwidth=-\maxdimen
}
\def\makeTABLE#1#2{
   {
   \let\ifmath0
   \let\header0
   \let\multispan0
%
%
   \ncase=0%
   \ifdim\tablewidth>-\maxdimen \ncase=1\fi%
   \ifdim\spreadwidth>-\maxdimen \ncase=2\fi%
   \relax
%
   \ifcase\ncase %
      \widthspec={}%
   \or %
      \widthspec=\expandafter{\expandafter t\expandafter o%
                 \the\tablewidth}%
   \else %
      \widthspec=\expandafter{\expandafter s\expandafter p\expandafter r%
                 \expandafter e\expandafter a\expandafter d%
                 \the\spreadwidth}%
   \fi %
   \xdef\next{
      \halign\the\widthspec{%
      #1
      \noalign{\hrule height\thicksize depth0pt}
      \the#2\endtable
%
      }
   }
   }
   \next
}
\def\makePREAMBLE#1{
   \ncols=#1
   \begingroup
   \let\ARGS=0
   \edef\xtp{\widevline\ARGS\tabskip\tabskipglue%
   &\ctr{\ARGS}\tstrut}
   \advance\ncols by -1
   \loop
      \ifnum\ncols>0 %
      \advance\ncols by -1%
      \edef\xtp{\xtp&\vrule width\thinsize\ARGS&\ctr{\ARGS}}%
   \repeat
   \xdef\preamble{\xtp&\widevline\ARGS\tabskip0pt%
   \crnorm}
   \endgroup
}
\def\countROWS#1\into#2{
   \let\countREGISTER=#2%
   \countREGISTER=0%
   \expandafter\ROWcount\the#1\endcount%
}%
\def\ROWcount{%
   \afterassignment\subROWcount\let\next= %
}%
\def\subROWcount{%
   \ifx\next\endcount %
      \let\next=\relax%
   \else%
      \ncase=0%
      \ifx\next\cr %
         \global\advance\countREGISTER by 1%
         \ncase=0%
      \fi%
      \ifx\next\endrow %
         \global\advance\countREGISTER by 1%
         \ncase=0%
      \fi%
      \ifx\next\crthick %
         \global\advance\countREGISTER by 1%
         \ncase=0%
      \fi%
      \ifx\next\crnorule %
         \global\advance\countREGISTER by 1%
         \ncase=0%
      \fi%
      \ifx\next\crthickneg %
         \global\advance\countREGISTER by 1%
         \ncase=0%
      \fi%
      \ifx\next\crnoruleneg %
         \global\advance\countREGISTER by 1%
         \ncase=0%
      \fi%
      \ifx\next\crneg %
         \global\advance\countREGISTER by 1%
         \ncase=0%
      \fi%
      \ifx\next\header %
         \ncase=1%
      \fi%
      \relax%
      \ifcase\ncase %
         \let\next\ROWcount%
      \or %
         \let\next\argROWskip%
      \else %
      \fi%
   \fi%
   \next%
}
\def\counthdROWS#1\into#2{%
\dvr{10}%
   \let\countREGISTER=#2%
   \countREGISTER=0%
\dvr{11}%
\dvr{13}%
   \expandafter\hdROWcount\the#1\endcount%
\dvr{12}%
}%
\def\hdROWcount{%
   \afterassignment\subhdROWcount\let\next= %
}%
\def\subhdROWcount{%
   \ifx\next\endcount %
      \let\next=\relax%
   \else%
      \ncase=0%
      \ifx\next\cr %
         \global\advance\countREGISTER by 1%
         \ncase=0%
      \fi%
      \ifx\next\endrow %
         \global\advance\countREGISTER by 1%
         \ncase=0%
      \fi%
      \ifx\next\crthick %
         \global\advance\countREGISTER by 1%
         \ncase=0%
      \fi%
      \ifx\next\crnorule %
         \global\advance\countREGISTER by 1%
         \ncase=0%
      \fi%
      \ifx\next\header %
         \ncase=1%
      \fi%
\relax%
      \ifcase\ncase %
         \let\next\hdROWcount%
      \or%
         \let\next\arghdROWskip%
      \else %
      \fi%
   \fi%
   \next%
}%
{\catcode`\|=13\letbartab
\gdef\countCOLS#1\into#2{%
   \let\countREGISTER=#2%
   \global\countREGISTER=0%
   \global\multispancount=0%
   \global\firstrowtrue
   \expandafter\COLcount\the#1\endcount%
   \global\advance\countREGISTER by 3%
   \global\advance\countREGISTER by -\multispancount
}%
\gdef\COLcount{%
   \afterassignment\subCOLcount\let\next= %
}%
{\catcode`\&=13%
\gdef\subCOLcount{%
   \ifx\next\endcount %
      \let\next=\relax%
   \else%
      \ncase=0%
      \iffirstrow
         \ifx\next& %
            \global\advance\countREGISTER by 2%
            \ncase=0%
         \fi%
         \ifx\next\span %
            \global\advance\countREGISTER by 1%
            \ncase=0%
         \fi%
         \ifx\next| %
            \global\advance\countREGISTER by 2%
            \ncase=0%
         \fi
         \ifx\next\|
            \global\advance\countREGISTER by 2%
            \ncase=0%
         \fi
         \ifx\next\multispan
            \ncase=1%
            \global\advance\multispancount by 1%
         \fi
         \ifx\next\header
            \ncase=2%
         \fi
         \ifx\next\cr       \global\firstrowfalse \fi
         \ifx\next\endrow   \global\firstrowfalse \fi
         \ifx\next\crthick  \global\firstrowfalse \fi
         \ifx\next\crnorule \global\firstrowfalse \fi
         \ifx\next\crnoruleneg \global\firstrowfalse \fi
         \ifx\next\crthickneg  \global\firstrowfalse \fi
         \ifx\next\crneg       \global\firstrowfalse \fi
      \fi
\relax
      \ifcase\ncase %
         \let\next\COLcount%
      \or %
         \let\next\spancount%
      \or %
         \let\next\argCOLskip%
      \else %
      \fi %
   \fi%
   \next%
}%
\gdef\argROWskip#1{%
   \let\next\ROWcount \next%
}
\gdef\arghdROWskip#1{%
   \let\next\ROWcount \next%
}
\gdef\argCOLskip#1{%
   \let\next\COLcount \next%
}
}
}
\def\spancount#1{
   \nspan=#1\multiply\nspan by 2\advance\nspan by -1%
   \global\advance \countREGISTER by \nspan
   \let\next\COLcount \next}%
\def\dvr#1{\relax}%
\def\header#1{%
\dvr{1}{\let\cr=\@mpersand%
\hdtks={#1}%
\counthdROWS\hdtks\into\hdrows%
\advance\hdrows by 1%
\ifnum\hdrows=0 \hdrows=1 \fi%
\dvr{5}\makehdPREAMBLE{\the\hdrows}%
\dvr{6}\getHDdimen{#1}%
{\parindent=0pt\hsize=\hdsize{\let\ifmath0%
\xdef\next{\valign{\headerpreamble #1\crnorm}}}\dvr{7}\next\dvr{8}%
}%
}\dvr{2}}
\def\makehdPREAMBLE#1{
\dvr{3}%
\hdrows=#1
{
\let\headerARGS=0%
\let\cr=\crnorm%
\edef\xtp{\vfil\hfil\hbox{\headerARGS}\hfil\vfil}%
\advance\hdrows by -1
\loop
\ifnum\hdrows>0%
\advance\hdrows by -1%
\edef\xtp{\xtp&\vfil\hfil\hbox{\headerARGS}\hfil\vfil}%
\repeat%
\xdef\headerpreamble{\xtp\crcr}%
}
\dvr{4}}
\def\getHDdimen#1{%
\hdsize=0pt%
\getsize#1\cr\end\cr%
}
\def\getsize#1\cr{%
\endsizefalse\savetks={#1}%
\expandafter\lookend\the\savetks\cr%
\relax \ifendsize \let\next\relax \else%
\setbox\hdbox=\hbox{#1}\newhdsize=1.0\wd\hdbox%
\ifdim\newhdsize>\hdsize \hdsize=\newhdsize \fi%
\let\next\getsize \fi%
\next%
}%
\def\lookend{\afterassignment\sublookend\let\looknext= }%
\def\sublookend{\relax%
\ifx\looknext\cr %
\let\looknext\relax \else %
   \relax
   \ifx\looknext\end \global\endsizetrue \fi%
   \let\looknext=\lookend%
    \fi \looknext%
}%
%
%
\def\tablelet#1{%
   \tableLETtokens=\expandafter{\the\tableLETtokens #1}%
}%
\catcode`\@=12

\def\({[}
\def\){]}
\def\l{{\lambda}}

\def\ea{{\epsilon_1}}
\def\eb{{\epsilon_2}}
\def\CH{{\cal H}}
\def\CL{{\cal L}}
\def\tM{{\tilde M}}
\def\tm{{\tilde m}}
\def\tV{{\tilde V}}

\def\ra{\rightarrow}
\def\mod{{\rm mod}}
\def\NP{Nucl. Phys. }
\def\PL{Phys. Lett. }
\def\PRL{Phys. Rev. Lett. }
\def\PRV{Phys. Rev. }

\def\PR{Phys. Rep. }
\def\lsim{{\roughly<}}
\def\gsim{{\roughly>}}

\def\psu{s^{\prime u}}
\def\psd{s^{\prime d}}

\Title{hep-ph/9310320, RU-93-43, WIS-93/93/Oct-PH}
{\vbox{\centerline{Mass Matrix Models: The Sequel}}}
\bigskip
\centerline{Miriam Leurer$^a$, Yosef Nir$^a$ and Nathan Seiberg$^b$}
\smallskip
\centerline{\it $^a$Department of Particle Physics}
\centerline{\it Weizmann Institute of Science, Rehovot 76100, Israel}
\smallskip
\centerline{\it $^b$Department of Physics and Astronomy}
\centerline{\it Rutgers University, Piscataway, NJ 08855-0849}
\bigskip
\baselineskip 18pt

\noindent
The smallness of the quark sector parameters and the hierarchy between
them could be the result of a horizontal symmetry broken by a small
parameter.  Such an explicitly broken symmetry can arise from an exact
symmetry which is spontaneously broken.  Constraints on the scales of
new physics arise {}from new flavor changing interactions and {}from
Landau poles, but do not exclude the possibility of observable
signatures at the TeV scale. Such a horizontal symmetry could also lead
to many interesting results: (i) quark -- squark alignment that would
suppress, without squark degeneracy, flavor changing neutral currents
induced by supersymmetric particles, (ii) exact relations between mass
ratios and mixing angles, (iii) a solution of the $\mu$-problem and (iv)
a natural mechanism for obtaining hierarchy among various symmetry
breaking scales.

\Date{10/93}

\newsec{Introduction}

Quark mass ratios and mixing angles have two intriguing features: The
{\it smallness} of most of these parameters and the {\it hierarchy}
among them. The hierarchy in the quark mixing angles is clearly presented
in the Wolfenstein's parameterization
\ref\wolf{L. Wolfenstein, \PRL\ {\bf 51} (1983) 1945.}\
of the CKM matrix:
\eqn\wolfenstein{V_{\rm CKM}=
\pmatrix{1-{\l^2\over 2}& \l & \l^3A(\rho+i\eta)\cr
-\l & 1- {\l^2\over 2} & \l^2A \cr
\l^3A(1-\rho+i\eta) & -\l^2A & 1 \cr} }
The hierarchy is reflected in the dependence of the various entries on
different powers of $\l\sim0.2$: all other quantities are experimentally
determined to be of order one, {\it i.e.} $A$ is of order 1 while $\rho$
and $\eta$ are between $\lambda$ and 1. The order of magnitude of the
three mixing angles is therefore given in powers of $\l$:
\eqn\mixlam{|V_{us}|\sim\l,\ \ |V_{cb}|\sim\l^2,\ \
|V_{ub}|\sim \l^3 - \l^4.}
The hierarchy among the quark masses can also be expressed in
powers of $\l$. The low energy ($\sim100\ GeV$) values are
\eqn\maslam{\eqalign{ m_u/m_c\sim & \l^3 - \l^4, \qquad
\qquad m_c/m_t\sim \l^3, \cr
m_d/m_s \sim & \l^2, \qquad
\qquad\qquad m_s/m_b \sim \l^2, \cr
m_b/m_t \sim & \l^2 -\l^3,
\qquad\qquad  m_t/\vev{\phi_u}\sim1. \cr}}
The exact power of $\l$ for each of the mass ratios may vary a little,
depending the top mass, and on the exact value one chooses for $\l$ in
the range $0.20-0.22$. Furthermore, all the parameters run under the
renormalization group and therefore, the order of magnitude estimates
\mixlam\ and \maslam\ depend on the scale.  Assuming that only the
top quark Yukawa coupling is large and using the one loop
renormalization group with the particle content of the minimal
supersymmetric standard model, we find that the only changes in our
order of magnitude
estimates at a high ($\sim10^{15}\ GeV$) scale are:
\eqn\maslamhigh{ m_c/m_t\sim \l^3 - \l^4, \qquad
\qquad m_b/m_t\sim \l^3.}

We would like to make two comments about the estimates in \maslam:
\item{(i)} It is possible that $m_u$ as deduced {}from first order
chiral Lagrangian predictions reflects non-perturbative strong
interaction effects rather than the value of the high energy parameter
\ref\georgi{H. Georgi and I.N. McArthur, Harvard preprint HUTP-81/A011
(1981), unpublished; D.B. Kaplan and A.V. Manohar, \PRL {\bf 56} (1986)
2004.}.
In particular, it could be that the bare $m_u$ vanishes, thus providing
a solution (which is natural in our framework
\ref\bns{T. Banks, Y. Nir and N. Seiberg, unpublished.})
to the strong CP problem. The phenomenological viability of this
scenario is controversial.
\item{(ii)} The small ratio ${m_b\over m_t}$ may be a result
of a large ratio of VEVs, $\tan\beta={\vev{\phi_u}\over\vev{\phi_d}}
\sim\l^{-2}-\l^{-3}$, or it may be the result of a small ratio of Yukawa
couplings when $\tan\beta\sim1$.

\noindent
For the large part of our study, these two points are not crucial, and
we use $m_u/m_c\sim\l^3$ and $\tan\beta\sim1$.

As articulated by 'tHooft
\ref\thooft{G. 'tHooft, Lecture at the Cargese Summer Institute,
(1979)},
small numbers are natural only if an exact symmetry is acquired when
they are set to zero (``naturalness'').  Therefore, both the smallness
of the quark sector parameters and the hierarchy among them may be
related to a symmetry --  a horizontal symmetry $\CH$ that acts on the
quarks (for recent discussions, see e.g.
\ref\hall{A. Antaramian, L.J. Hall and A. Rasin, \PRL {\bf 69} (1992)
1871; L.J. Hall and S. Weinberg,  \PRV {\bf D48} (1993) R979.}).
Such a horizontal symmetry may be responsible for the hierarchy, if
it is explicitly broken by an operator in the Lagrangian whose
coefficient is the small parameter $\l$. The transformation laws of $\l$
under $\CH$ control the order in perturbation theory of the various
elements in the quark mass matrices and, consequently, some parameters
depend on powers of $\l$ higher than others, namely a hierarchy can be
generated. This phenomenon is common in atomic physics, nuclear physics
and particle physics and is known as ``selection rules.''

The next step is to understand the origin of this {\it explicitly
broken} $\CH$.  Several different mechanisms can exist.  Here we focus
on the possibility that $\l$ is promoted to a quantum field which has an
expectation value.  This expectation value {\it spontaneously breaks}
the exact symmetry $\CH$.  More precisely, we add a scalar field $S$
whose expectation value $\vev{S}$ breaks $\CH$.  The small numbers in
the Lagrangian appear then as powers of the ratio $\l={\vev{S}\over M}$
where $M$ is a higher energy scale at which the information about
$\CH$-breaking is communicated to the light fermions.

The organization of this paper follows this logic.  First in sections 2
and 3 we consider an explicitly broken $\CH$ and examine its
consequences.  The simplest framework, that of an Abelian group\foot{We
do not study non-Abelian symmetries. An example of a non-Abelian model
is presented in reference
\ref\philippe{P. Pouliot and N. Seiberg, Rutgers preprint RU-93-39
(1993), hep-ph/9308363, \PL in press.}.},
$\CH\subset U(1)$, and a single small breaking parameter $\l$, is
described in subsection 2.1. There is an essentially unique model in this
framework.  The next-to-simplest symmetry, $\CH\subset U(1)\times
U(1)$ with two small breaking parameters (one for each $U(1)$) opens
up interesting possibilities: First, the hierarchy in the quark
parameters can be achieved with lower powers of the small parameters.
This is demonstrated in an example in subsection 2.2, that is the basis
for an interesting high-energy model to be presented later. Second, we
can acquire highly suppressed entries in the quark mass matrices. This
may lead to suppression of FCNC induced by squark-gluino diagrams. We
explain this mechanism (`Quark -- Squark Alignment')
\ref\qsa{Y. Nir and N. Seiberg, \PL {\bf B309} (1993) 337.}
in subsection 2.3. Phenomenological constraints are discussed in
subsection 2.4.

The possibility of acquiring zero or highly suppressed entries in
the quark mass matrices allows relations that go beyond the
naive order of magnitude estimates. The possibility that $|V_{ub}| \ll
|V_{us}V_{cb}|$ (rather than of the same order of magnitude) is
discussed in subsection 3.1, while close to exact relations between
mixing angles and mass ratios are obtained in subsection 3.2.

Sections 4 --- 6 are concerned with the embedding of the low energy
effective theory with explicitly broken horizontal symmetry in a more
fundamental theory. As mentioned above, the most natural theory takes
$\CH$ to be an exact symmetry, spontaneously broken by the VEVs of
scalar fields. The mechanism, its phenomenological consequences and
constraints on the scale of spontaneous symmetry breaking are described
in section 4. A very plausible explanation of the physics at the scale
$M$, where the information about the spontaneous breaking is
communicated to the light fermions, was suggested by Froggatt and
Nielsen
\ref\frni{C.D. Froggatt and H.B. Nielsen,
 Nucl. Phys. {\bf B147} (1979) 277.}
and was further studied in references
\ref\dimo{S. Dimopoulos, \PL {\bf 129B} (1983) 417;
J. Bagger, S. Dimopoulos, E. Masso and M. Reno, \NP {\bf B258} (1985)
565;
J. Bagger, S. Dimopoulos, H. Georgi and S.~Raby, In: {\it Proc. Fifth
Workshop on Grand Unification.} Eds. Kang, K., Fried, H. and Frampton,
P., Singapore, World Scientific (1984);
Z.G. Berezhiani, \PL {\bf B129} (1983) 99, {\bf B150} (1985) 177;
A. Davidson, V.P. Nair and K.C. Wali, \PRV {\bf D29} (1984) 1505;
A. Davidson and K.C. Wali, \PRL {\bf 60} (1988) 1313;
A. Davidson, S. Ranfone and K.C. Wali, \PRV {\bf D41} (1990) 208.}
and
\ref\lns{M. Leurer, Y. Nir and N. Seiberg, \NP {\bf B398} (1993) 319.}:
$M$ is the mass scale for heavy mirror quarks. The scalars responsible
for the breaking couple the heavy sector to the light one.  This
mechanism, its phenomenological consequences and constraints on the
scale $M$ are studied in section 5. The existence of extra
supermultiplets affects the running of the various gauge couplings. In
section 6 we investigate the constraints that follow if we assume that
there is no further new physics between $M$ and the Planck scale $M_P$
and require the absence of Landau poles.  This analysis allows us to
address the important question of whether flavor physics may be directly
observed in future experiments.

Our full framework requires that there are several new scales of physics
between the SUSY breaking scale and the Planck scale.  In section 7 we
study the $\CH$-invariant Higgs potential and suggest a mechanism that
can naturally generate the required hierarchy. Furthermore, this
mechanism automatically solves the ``$\mu$-problem.'' A discussion of
our results is given in section 8.

Our discussion proceeds {}from low energies to high energies.  At every
step describing the physics at higher energies we add more assumptions
which lead to more constraints.  It is possible that our ideas about
some energy scale will turn out to be correct while the speculations
about higher energies will turn out to be wrong.

\newsec{Models of Explicitly Broken Horizontal Symmetry}

\subsec{The Master Model}

The simplest framework
to explain the order of magnitude relations
\mixlam\ and \maslam\ is that of a horizontal symmetry $\CH=U(1)_H$,
with a small breaking parameter $\l\sim0.2$. Terms that break $\CH$ by
$n(>0)$ units of charge are suppressed by $\l^n$. In the full
Lagrangian, we require that $\CH$ is a discrete subgroup, $Z_N\subset
U(1)$. Then terms that break $\CH$ by $n>N$ units of charge are
suppressed only by $\l^l$ for $l=n~\mod ~ N$. In most of the cases that
we are interested in, $N$ is large enough and this does not happen.
Therefore, for convenience, we will treat $\CH$ as a continuous $U(1)$
symmetry. In the few cases where the discreteness of the symmetry does
play an important role, we explicitly point out its effects.

We will usually restrict ourselves to supersymmetric theories (with
supersymmetry broken by soft terms). Then terms in the fermion mass
matrices that break the horizontal symmetry by $n<0$ units of charge
vanish. This is due to our assumption that $\l$ is a (single)
coefficient of an operator which explicitly breaks $\CH$ and the fact
that the effective superpotential is holomorphic in the coupling
constants of the theory (no powers of $\l^\dagger$).  This is a special
case of the non-renormalization theorem of reference
\ref\sei{N. Seiberg, Rutgers preprint RU-93-45 (1993), hep-ph/9309335.}.
If more than one small $\l$ which break the same symmetry exist, say one
with negative charge and another with positive charge, then such terms
need not vanish.  The consequences of this fact will be discussed below.

We use the following notation for the various fields and their
interactions. $Q_i$ denote the left-handed quark doublets and $\bar d_i$
($\bar u_i$) denote the left handed anti-down (anti-up) $SU(2)$
singlets. $\phi_d$ and $\phi_u$ are the Higgs fields of hypercharge
$-1/2$ and $+1/2$, respectively. The Yukawa couplings are denoted by
$Y^d$ and $Y^u$, and the mass matrices by $M^d$ and $M^u$.  The
Yukawa interactions are then given by
\eqn\yukmssm{\CL_Y=Y^d_{ij}\phi_d Q_i\bar d_j
+Y^u_{ij}\phi_u Q_i\bar u_j.}

The interaction \yukmssm\ has an accidental $U(1)$ symmetry, which we
call $U(1)_X$. Under this symmetry $\phi_d$ carries charge $-1$, the
$\bar d_i$ fields carry charge $+1$, while all other fields carry
vanishing
$X$ charges. The $U(1)_X$ symmetry must be explicitly broken in other
sectors of the Lagrangian, since otherwise its spontaneous breaking at
the weak scale implies the existence of an unwanted axion. Despite being
broken, $U(1)_X$ turns out to be extremely useful. First, using
$U(1)_X$, hypercharge and baryon number symmetries, we can set the
horizontal charges of $\phi_u$, $\phi_d$ and $Q_3$ to zero. As long as
we restrict our discussion to $U(1)_X$ invariant terms in the
Lagrangian, such a horizontal charge redefinition is justified and
simplifies the analysis considerably. Second, the existence of $U(1)_X$
in the Yukawa sector implies that QCD anomalies pose no problem to our
horizontal symmetries. It is always possible to find a horizontal
symmetry $\tilde\CH\subset \CH\times U(1)_X$ that restricts the quark
mass matrices in precisely the same way as does $\CH$ but is free of QCD
anomaly. We discuss this point in detail in subsection 4.3.

The order of magnitude of the Yukawa couplings is determined by their
horizontal quantum numbers. Assuming positive (or vanishing) horizontal
charges to all quark fields one finds \frni:
\eqn\suppress{Y^d_{ij}\sim\l^{H(Q_i)+H(\bar d_j)},\ \ \
Y^u_{ij}\sim\l^{H(Q_i)+H(\bar u_j)}.}
The mixing angles and mass ratios can then be estimated:
\eqn\mastermix{|V_{ij}|\sim\ \l^{|H(Q_i)-H(Q_j)|},}
\eqn\mastermas{{m_{d_i}\over m_{d_j}}\sim\l^{H(Q_i)-H(Q_j)+
H(\bar d_i)-H(\bar d_j)},\ \ \
{m_{u_i}\over m_{u_j}}\sim\l^{H(Q_i)-H(Q_j)+
H(\bar u_i)-H(\bar u_j)}.}
(Both \mastermix\ and \mastermas\ are given here for $i<j$.)

The order of magnitude estimates \mixlam\ and \maslam\ then
determine a {\it unique} set of $H$-charges for all quark fields
(when we take\foot{It is trivial to modify the charges for the other
possibilities.} $m_u/m_c\sim\l^3$, $m_b/m_t\sim\l^2$, $\tan\beta\sim1$):
\eqn\mastercharge{\matrix{Q_1&Q_2&Q_3&&\bar d_1&\bar d_2&\bar d_3&&
\bar u_1&\bar u_2&\bar u_3\cr
(3)&(2)&(0)&&(3)&(2)&(2)&&(3)&(1)&(0)\cr}}
With these charges the mass matrices have the order of magnitude entries
\eqn\mastermass{
M^d\sim\vev{\phi_d}\pmatrix{\l^6&\l^5&\l^5\cr \l^5&\l^4&\l^4\cr
\l^3&\l^2&\l^2\cr},\ \ \
M^u\sim\vev{\phi_u}\pmatrix{\l^6&\l^4&\l^3\cr \l^5&\l^3&\l^2\cr
\l^3&\l&1\cr}.}
It is straightforward to check that these mass matrices indeed lead
to mixing angles and mass ratios as given in \mixlam\ and \maslam.
We also note that the determinants of the mass matrices are
\eqn\masterdet{\det M^d\sim\vev{\phi_d}^3\l^{12},\ \ \
\det M^u\sim\vev{\phi_u}^3\l^9.}
The powers of $\l$ in these determinants will be of importance in
the discussion of the embedding of the model in a high energy theory.

How predictive is this framework? We have made eight discrete choices of
charges and explained the order of magnitude of nine physical
(continuous) parameters ($m_t/\vev{\phi_u}\sim1$, $m_b/m_t\sim \l^2$,
the four mass ratios of eq. \mastermas, and the three mixing angles of
equation \mastermix). This means that the model predicts a single
order of magnitude relation. Indeed, we find
\eqn\ubcbus{|V_{ub}/V_{cb}|\sim |V_{us}|}
independent of the choice of charges. The Particle Data Group quotes
\ref\pdg{K. Hikasa {\it et al.}, Particle Data Group, \PRV {\bf D45}
 (1992) S1.}
\eqn\pdgub{|V_{ub}/V_{cb}|=0.10\pm0.03,\ \ \ |V_{us}|=0.22,}
consistent with \ubcbus. Recently, however, the CLEO collaboration
announced a new measurement
\ref\cleovub{I.P.J. Shipsey (CLEO Collaboration), a talk given in the
International Europhysics Conference on High Energy Physics,
Marseille 1993.}:
\eqn\cleoub{|V_{ub}/V_{cb}|\sim\ 0.05\ -\ 0.10.}
We remind the reader that there is a strong theoretical model dependence
in the extraction of $|V_{ub}/V_{cb}|$ {}from the experimental data.
However, if it eventually turns out that $|V_{ub}/V_{cb}|\sim\l^2$,
it will pose a problem to the naive master model presented here.
In subsection 3.1 we show how the relation \ubcbus\ can be avoided
in a more sophisticated model within our framework.

We take $\CH$ to commute with Supersymmetry. The alternative, that
$\CH$ is an R symmetry will be briefly discussed below. Then the
charges \mastercharge\ are common to complete supermultiplets.
This determines the form of the squark mass-squared matrices as well.
We denote these by $\tM^{d2}$ and $\tM^{u2}$,
and divide each to $3\times3$ sub-matrices:
\eqn\tMblocks{\tM^{d2}=\pmatrix{\tM^{d2}_{LL} & \tM^{d2}_{LR} \cr
(\tM^{d2}_{LR})^\dagger & \tM^{d2}_{RR} \cr},\ \ \
\tM^{u2}=\pmatrix{\tM^{u2}_{LL} & \tM^{u2}_{LR} \cr
(\tM^{u2}_{LR})^\dagger & \tM^{u2}_{RR} \cr}.}
The sub-index $L$ ($R$) refers to the scalar partners of the quark
doublets $Q_i$ (singlets $\bar d_i$ or $\bar u_i$). The
leading contributions to the diagonal blocks
arise {}from $A$-type SUSY breaking terms, while
the leading contributions to the off diagonal blocks arise {}from soft
SUSY
breaking terms analytical in the fields. This leads to two
important differences between the diagonal and the off-diagonal blocks:
(i) Entries in the diagonal blocks that break $\CH$ by $n$ units of
charge are suppressed by $\l^{|n|}$, whether $n$ is negative or positive.
In the off-diagonal blocks such entries are suppressed by $\l^n$ for
$n>0$, but vanish when $n<0$. (ii) The entries in the diagonal blocks
are proportional to $\tilde m^2$ ($\tilde m$ is the scale of SUSY
breaking) while those in the off-diagonal blocks
are proportional to $\tilde m \vev{\phi_q}$. If $\tm\gg\vev{\phi_{q}}$,
the off diagonal blocks are negligible to all our purposes.

The charge assignments \mastercharge\ give
\eqn\mastertmll{\tM^{d2}_{LL}\approx \tM^{u2}_{LL}\sim\tm^2\pmatrix
{1 & \l & \l^3 \cr \l & 1 & \l^2 \cr \l^3 & \l^2 & 1 \cr},}
\eqn\mastertmrr{\tM^{d2}_{RR}\sim\tm^2\pmatrix
{1 & \l & \l \cr \l & 1 & 1 \cr \l & 1 & 1 \cr},\ \ \
\tM^{u2}_{RR}\sim\tm^2\pmatrix
{1 & \l^2 & \l^3 \cr \l^2 & 1 & \l \cr \l^3 & \l & 1 \cr},}
\eqn\mastertmlr{(\tM^{d2}_{LR})_{ij}\sim\tm M^d_{ij},\ \ \
(\tM^{u2}_{LR})_{ij}\sim\tm M^u_{ij}.}
We remind the reader that all entries are order of magnitude estimates
and not exact numbers. (In particular, the diagonal elements in each of
$\tM^{d2}$ and $\tM^{u2}$ are all of order $\tm^2$ but not equal to
each other.) However,
for $\tm$ considerably higher than the electroweak breaking scale,
the approximate equality $\tM^{d2}_{LL}\approx\tM^{u2}_{LL}$
holds to $\CO(\vev{\phi}^2/\tm^2)$ and not just as
an order of magnitude estimate.

If $\CH$ is an $R$ symmetry, the diagonal blocks $\tM^{d2}_{LL}$,
$\tM^{d2}_{RR}$, $\tM^{u2}_{LL}$ and $\tM^{u2}_{RR}$ are unchanged.
The off-diagonal blocks are different and, unlike eq. \mastertmlr,
their suppression by powers of $\l$ is not the same as for the
corresponding elements in the quark mass matrices. Then their
effects on FCNC can be significant (even though they are suppressed by
$\vev{\phi}/\tm$). Thus, our discussion of quark -- squark alignment
in subsection 2.3 does not apply in general to horizontal R symmetries.

\subsec{Models with $U(1)_{H_1}\times U(1)_{H_2}$ Symmetry}

Models with a more complicated symmetry structure than a simple $U(1)$
offer new possibilities in constructing mass matrices. All the important
features of such symmetries are already present in the simplest
extension,
\eqn\doubleuone{\CH=U(1)_{H_1}\times U(1)_{H_2},}
with two small breaking parameters:
\eqn\lalb{\ea\sim\l^p,\ \ \ \eb\sim\l^q,}
where $q>p\ge 1$.

Let us compare a model with the horizontal symmetry \doubleuone\
to the master model of the previous section. A quark field that
carries charge $H$ in the master model, must carry charges
$(H_1,H_2)$ under \doubleuone\ such that
\eqn\abcharges{H=p H_1+q H_2.}
We note the following points:
\item{(i)} The choice of $(H_1,H_2)$ is, in general, not unique.  Thus,
unlike the master model, for each horizontal symmetry of the type
\doubleuone, there are several models (namely, sets of charge
assignments for the quark fields) that produce the same hierarchy in
mixing angles and mass ratios.
\item{(ii)} Consider the determinants of the mass matrices. For example,
\eqn\detmd{\det M^d=\ea^{\sum_i(H_1(Q_i)+H_1(\bar d_i))}
               \eb^{\sum_i(H_2(Q_i)+H_2(\bar d_i))},}
to be compared with the master model
\eqn\detmdmr{\det M^d({\rm master})=\l^{\sum_i(H(Q_i)+H(\bar d_i))}.}
Since $q>1$, it is trivial to show that the sum of powers of the
$\epsilon_i$'s in the new model is smaller than the power of
$\l$ in the master model. When we later study the
underlying theory, we will find that the lower the power of $\epsilon$,
the weaker is the lower bound on the horizontal physics scale.
\item{(iii)} In the master model, all entries in the mass matrices have
their naively expected values. In the model of eq. \doubleuone,
some entries are suppressed and would vanish if the horizontal symmetry
were continuous. For example, $M^d_{ij}$ would vanish if either
$H_1(Q_i)+H_1(\bar d_j)<0$ or $H_2(Q_i)+H_2(\bar d_j)<0$.
Such suppressed entries open interesting possibilities.
In particular, we find that it is possible to solve the problem
of the squark-gluino box diagram contribution to neutral meson mixing
{\it without} requiring squark degeneracy (see next subsection).
Another interesting consequence is the possibility of relations between
mixing angles and mass ratios (see section 3).

Let us present an explicit example. The symmetry is of the type
\doubleuone\ with breaking parameters
\eqn\modela{\ea\sim\l^2,\ \ \ \eb\sim\l^3.}
(To explain $|V_{us}|\sim\l$, we always need either $p=1$ or $q-p=1$.)
There are four models that reproduce the order of magnitude
relations of the master model. This is a result of two possible choices
for the charges of each of $\bar d_1$ and $\bar u_1$: $(0,1)$ or
$(3,-1)$. In one of the four models the charges are:
\eqn\modelacharge{\matrix{Q_1&Q_2&Q_3&&\bar d_1&\bar d_2&\bar d_3&&
\bar u_1&\bar u_2&\bar u_3\cr
(0,1)&(1,0)&(0,0)&&(3,-1)&(1,0)&(1,0)&&(0,1)&(-1,1)&(0,0)\cr}}
The mass matrices have the order of magnitude entries
\eqn\modelamass{
M^d\sim\vev{\phi_d}\pmatrix{
\ea^3&\ea\eb&\ea\eb\cr 0&\ea^2&\ea^2\cr 0&\ea&\ea\cr},\ \ \
M^u\sim\vev{\phi_u}\pmatrix{\eb^2&0&\eb\cr \ea\eb&\eb&\ea\cr
\eb&0&1\cr}.}
We see that each entry in the mass matrix is either of the same order of
magnitude as in the master model, or zero. When we take into account the
fact that $\CH$ is discrete and not continuous, we find that the
vanishing entries are modified but still very suppressed relative to
their values in the master model. It is again straightforward to check
that these mass matrices
lead to mixing angles and mass ratios as given in \mixlam\ and \maslam.

The order of magnitude estimates of the squark mass-squared matrices
in this model are:
\eqn\mastertmll{\tM^{d2}_{LL}\approx\tM^{u2}_{LL}\sim\tm^2\pmatrix
{1 & \ea\eb & \eb \cr \ea\eb & 1 & \ea \cr \eb & \ea & 1 \cr},}
\eqn\mastertmrr{\tM^{d2}_{RR}\sim\tm^2\pmatrix
{1 & \ea^2\eb & \ea^2\eb \cr
\ea^2\eb &1&1\cr \ea^2\eb & 1 & 1 \cr},\ \ \
\tM^{u2}_{RR}\sim\tm^2\pmatrix
{1 & \ea & \eb \cr \ea & 1 & \ea\eb \cr \eb & \ea\eb & 1 \cr},}
\eqn\modelatmlr{(\tM^{d2}_{LR})_{ij}\sim\tm M^d_{ij},\ \ \
(\tM^{u2}_{LR})_{ij}\sim\tm M^u_{ij}.}

\subsec{Quark Squark Alignment}

For generic squark masses, box diagrams with squarks and gluinos give
unacceptably large contributions to neutral meson ($K$, $B$ and $D$)
mixing
\ref\squade{R. Barbieri and R. Gatto, \PL {\bf 110B} (1982) 211;
J. Ellis and D.V. Nanopoulos, \PL {\bf 110B} (1982) 44; H.P. Nilles, \PR
{\bf 110} (1984) 1; F. Gabbiani and A. Masiero, \NP {\bf B322} (1989)
235.}.
The standard solution to this problem is to assume that squarks
are degenerate to a very good approximation. This is not motivated in
generic supergravity models or string theory, though it may hold under
special conditions
\ref\vadim{V.S. Kaplunovsky and J. Louis, \PL {\bf B306} (1993) 269;
R. Barbieri, J. Louis and M. Moretti, \PL {\bf B312} (1993) 451.}.
Both squark degeneracy and proportionality of trilinear Higgs--squark
couplings to Yukawa couplings can be natural if supersymmetry breaking
is communicated to the light particles by gauge interactions
\ref\oldnewdine{M. Dine, A. Kagan and S. Samuel, \PL {\bf B243} (1990)
250; M. Dine and A. Nelson, \PRV {\bf D48} (1993) 1277.},
or in models with a non-Abelian horizontal symmetry
\ref\dkl{M. Dine, A. Kagan and R. Leigh, SCIPP-93/04 (1993),
hep-ph/9304299.}
\philippe.

In reference \qsa\ an alternative mechanism to suppress squark
contributions to FCNC was suggested: the approximate alignment of quark
mass matrices with squark mass-squared matrices.  The idea is that a
horizontal symmetry, of the type discussed in this work, forces both
$M^q$ and $\tM^{q2}$ to be approximately diagonal in the basis where the
horizontal charges are well defined.  Consequently, the mixing matrix
for quark -- squark -- gluino couplings is close to a unit matrix and
FCNC are suppressed, regardless of whether squarks are degenerate or
not.

To make the discussion concrete, we define the diagonalizing matrices
for quarks,
\eqn\vquark{\eqalign{
V_L^d M^d V_R^{d\dagger}=&{\rm diag}(m_d,m_s,m_b),\cr
V_L^u M^u V_R^{u\dagger}=&{\rm diag}(m_u,m_c,m_t),\cr}}
for down squarks,
\eqn\tvsquark{\eqalign{
\tV_L^d \tM^{d2}_{LL}\tV_L^{d\dagger}=&
{\rm diag}(m^2_{\tilde d_L},m^2_{\tilde s_L},m^2_{\tilde b_L}),\cr
\tV_R^d \tM^{d2}_{RR}\tV_R^{d\dagger}=&
{\rm diag}(m^2_{\tilde d_R},m^2_{\tilde s_R},m^2_{\tilde b_R}),\cr}}
and similarly for up squarks. Here we assume $\tm \gg \vev{\phi_{u,d}}$.
Then, the CKM matrix is $V=V_L^u V_L^{d\dagger}$, while the mixing
matrices for gluino interactions are
\eqn\mixgluino{\eqalign{
K_L^d=V_L^d\tV_L^{d\dagger},&\ \ \ K_R^d=V_R^d\tV_R^{d\dagger},\cr
K_L^u=V_L^u\tV_L^{u\dagger},&\ \ \ K_R^u=V_R^u\tV_R^{u\dagger}.\cr}}

Various FCNC processes, and in particular the mixing of neutral
mesons, put upper bounds on elements of the $K_M^q$ matrices
($M=L,R$; $q=d,u$). The bounds are particularly strong on
\eqn\kvev{\vev{K^q_{ij}}=\sqrt{(K^q_L)_{ij}(K^q_R)_{ij}}.}
For $m_{\tilde g}=\tilde m=1\ TeV$, the bounds are:\foot{
Usually, these bounds are applied to
the off diagonal entries in the squark mass matrices in the
basis where the quark mass matrices are diagonal. When the squarks
are not even approximately degenerate, as is the case in our discussion,
the bounds are on the matrix elements of $K$.}
\eqn\kconstraint{\eqalign{
\Delta m_K\ \Longrightarrow&\ (K_M^d)_{12}\lsim0.05,\ \
 \vev{K^d_{12}}\lsim0.006,\cr
\epsilon_K\ \Longrightarrow&\ (K_M^d)_{12}\lsim0.004,\ \
 \vev{K^d_{12}}\lsim0.0005,\cr}}
\eqn\dbconstraint{\eqalign{
\Delta m_D\ \Longrightarrow&\ (K_M^u)_{12}\lsim0.1,\ \
 \vev{K^u_{12}}\lsim0.04,\cr
\Delta m_B\ \Longrightarrow&\ (K_M^d)_{13}\lsim0.1,\ \
 \vev{K^d_{13}}\lsim0.04.\cr}}
 A few points are in order regarding these constraints:
\item{(i)} There are also bounds on the mixing matrices
$K_{LR}^q=V_L^q\tV_R^{q\dagger}$ and
$K_{RL}^q=V_R^q\tV_L^{q\dagger}$. However, these bounds are
easily satisfied in our framework \qsa\ and we do not present them here.
\item{(ii)} The bound on $\epsilon_K$ is valid only when we assume
that all CP violating phases are arbitrary and of order one.
\item{(iii)}  We emphasize that
there is an ambiguity of a factor of a few in these bounds, coming
{}from the exact value of the $\tm$ scale; {}from possible differences
between the gluino mass $m_{\tilde g}$ and the average squark mass
$\tilde m$; and {}from hadronic uncertainties in matrix elements of
quark operators.

Generically, the horizontal symmetries employed in our various models
give
\eqn\generick{\eqalign{
(K_L^d)_{12}\lsim\l\ ,\ \ (K_R^d)_{12}\lsim\l\ ,\ \
\vev{K^d}_{12}\lsim\l\ &,\cr
(K_L^u)_{12}\lsim\l\ ,\ \ (K_R^u)_{12}\lsim\l^2,\ \
\vev{K^u}_{12}\lsim\l^{3\over2}&,\cr
(K_L^d)_{13}\lsim\l^3,\ \ (K_R^d)_{13}\lsim\l\ ,\ \
\vev{K^d}_{13}\lsim\l^2&.\cr
}}
This means that the main problem is the suppression of the squark
contributions to $\Delta m_K$ and $\epsilon_K$; the contributions
to $\Delta m_D$ and $\Delta m_B$ are generically suppressed to just
an acceptable level. We now describe a class of models that we call
``Quark-Squark-Alignment" (QSA) models, in which
squark contributions to $\Delta M_K$ and to $\epsilon_K$ are
highly suppressed and \kconstraint\ is satisfied.

We again focus on models with $\tan\beta\sim1$ and $m_b/m_t\sim\l^2$,
but models with satisfactory quark--squark alignment exist also for
$\tan\beta\sim m_t/m_b$ or for $m_b/m_t\sim\l^3$.  The main problem is
to avoid $(V_L^d)_{12}\sim\l$ and $(V_R^d)_{12}\sim\l$, while keeping
the CKM values $|V_{us}|\sim\l$ and $|V_{ub}|\sim\l^3$. The expressions
for the elements of the diagonalizing matrices in terms of elements of
the mass matrices are given in Appendix A. Using these expressions, we
find that, to satisfy \kconstraint, the following entries in $M^d$ have
to vanish: $M^d_{12}$, $M^d_{21}$, either $M^d_{13}$ or $M^d_{32}$ and
either $M^d_{31}$ or $M^d_{23}$.  We stress again that when we say that
a particular $M^d_{ij}$ vanishes, we actually mean that it would vanish
if $\CH$ were continuous.  However, as $\CH$ is discrete, $M^d_{ij}$ is
not zero but only highly suppressed compared to its value in
\mastermass.

It is impossible to get vanishing (or suppressed) entries in models with
$\CH=U(1)$. We therefore scanned models with $\CH=U(1)\times U(1)$.
Some models with $\ea=\l^2$ and $\eb=\l^3$ give satisfactory
quark-squark alignment (such a model was presented in ref. \qsa), but
run into other phenomenological problems. Specifically, $|V_{td}|$ is
highly suppressed, leading to an unacceptably small $B-\bar B$ mixing.
On the other hand, we found many acceptable models with $\ea=\l$ and
$\eb=\l^2$ that lead to satisfactory suppression of the squark
contributions to $\Delta m_K$ and $\epsilon_K$.  The down quarks mass
matrix in these models is always of the form
\eqn\mdqsa{M^d\sim\vev{\phi_d}\pmatrix{\l^4&0&\l^3\cr 0&\l^2&\l^2\cr
0&0&1\cr},}
which leads to the following order of magnitude estimates:
\eqn\vonetwoqsa{(V_L^d)_{12}\sim\l^5,\ \ \ (V_R^d)_{12}\sim\l^7.}
Let us present one explicit example, which has also interesting
implications for CP asymmetries in $B$ decays:
\eqn\qsacharge{\matrix{Q_1&Q_2&Q_3&&\bar d_1&\bar d_2&\bar d_3\cr
(3,0)&(0,1)&(0,0)&&(-1,2)&(4,-1)&(0,1).\cr}}
For the down squark masses, we find in this model
\eqn\modelcmll{\tM^{d2}_{LL}\sim\tm^2\pmatrix
{1 & \ea^3\eb & \ea^3 \cr \ea^3\eb & 1 & \eb \cr \ea^3 & \eb & 1 \cr},}
\eqn\modelcmrr{\tM^{d2}_{RR}\sim\tm^2\pmatrix
{1 & \ea^5\eb^3 & \ea\eb \cr
\ea^5\eb^3 & 1 & \ea^4\eb^2 \cr \ea\eb & \ea^4\eb^2 & 1 \cr},}
\eqn\modelcmlr{(\tM^{d2}_{LR})_{ij}\sim\tm M^d_{ij}.}
This gives
\eqn\vtildeqsa{(\tV_L^d)_{12}\sim\l^5,\ \ \ (\tV_R^d)_{12}\sim\l^{11},}
which, together with eq. \vonetwoqsa, leads to
\eqn\modelck{(K_L^d)_{12}\sim\l^5,\ \ (K_R^d)_{12}\sim\l^7,}
consistent with \kconstraint.

Let us now see how the constraint on $\Delta m_D$ in eq.
\dbconstraint\ is satisfied in this model. For that, we have to
assign horizontal charges to the $\bar u_i$ fields. Take as an example
\eqn\ucharge{\matrix{\bar u_1&\bar u_2&\bar u_3\cr
(-1,2)&(1,0)&(0,0)\cr}.}
The constraints on $(K^u_R)_{12}$ and $\vev{K^u}_{12}$ are easily
satisfied, but the constraint on $(K^u_L)_{12}$ is only barely so.
Then the model predicts that $\Delta m_D$ is very close to the
experimental upper bound. This is actually not just a feature of the
model presented here, but a crucial test of the quark--squark alignment
idea: in all QSA models, $(V_L^d)_{12}$ is highly suppressed and,
therefore, $(V_L^u)_{12}$ must be equal to the Cabibbo angle,
namely $(V_L^u)_{12}\sim\l$. This gives
\eqn\qsakuonetwo{(K^u_L)_{12}\sim\l,}
which is at the order of the upper bound. The conclusion is that in all
QSA models, $D-\bar D$ mixing is orders of magnitude above the Standard
Model and should be very close to its present upper bound.

The QSA model presented here has also interesting implications for
$B$-physics. A rough estimate of the ratio between the SUSY contribution
to $B-\bar B$ mixing and the Standard Model one gives
\eqn\bsusysm{
{|M_{12}(B^0)|_{\rm SUSY}\over|M_{12}(B^0)|_{\rm SM}}\approx
250\((K^d_L)_{13}^2+(K^d_R)_{13}^2\)+2500\vev{K^d_{13}}^2,}
where we used vacuum-insertion approximation for the various matrix
elements and a scale $\tm\sim1\ TeV$.  {}From equations \mdqsa,
\modelcmll, \modelcmrr\ and \modelcmlr\ we find
\eqn\modelcvub{(K_L^d)_{13}\sim\l^3,\ \ \ (K_R^d)_{13}\sim\l^3,}
which gives ${|M_{12}(B^0)|_{\rm SUSY}\over |M_{12}(B^0)|_{\rm SM}}
\approx 0.15$. While this contribution is small enough to satisfy the
$\Delta m_B$ constraint in \dbconstraint,
it may lead to observable effects in
CP asymmetries in $B$ decays. It is important here that the
quark--squark alignment is precise enough to satisfy the $\epsilon_K$
constraint: this means that we have no reason to assume that the new CP
violating phases in the $K_M^q$ matrices are small. With new phases of
$\CO(1)$, and with magnitude which is $\CO(0.15)$ of the Standard Model
one, the shift {}from the Standard Model predictions in CP asymmetries
in the decays of neutral $B$ into final CP eigenstates may be as large
as $\CO(0.3)$.

The potentially large effect on CP asymmetries in $B^0$ decays is not a
generic feature of quark -- squark alignment models.  Actually, it is
possible to show that, while $(K^d_L)_{13}\sim\l^3$ in all our models of
quark -- squark alignment, the order of magnitude estimate \modelcvub\
is an {\it upper bound} on $(K_R^d)_{13}$ in this framework. The fact
that the specific model presented in this subsection saturates this
bound and, therefore, gives interesting effects was our reason to
present it in the first place.  In most other models
$(K_R^d)_{13}\sim\l^5$ or even $\l^7$, so that the shift {}from the
Standard Model predictions for CP asymmetries is of $\CO(0.01)$ or even
less.  A measurement of the asymmetries may then distinguish among our
various models.

A similar investigation can be made for CP asymmetries in $B_s$ decays:
\eqn\bssusysm{
{|M_{12}(B_s)|_{\rm SUSY}\over|M_{12}(B_s)|_{\rm SM}}\approx
10\((K^d_L)_{23}^2+(K^d_R)_{23}^2\)+100\vev{K^d_{23}}^2.}
For models of quark -- squark alignment
\eqn\modelcvcb{(K_L^d)_{23}\sim\l^2,\ \ \ (K_R^d)_{13}\lsim\l^4.}
The shift {}from the Standard Model predictions for CP asymmetries in
$B_s$ decays is then of $\CO(0.01)$. This is probably too small to be
experimentally observed. It also leads to the interesting situation
\ref\nisi{Y. Nir and D. Silverman, \NP {\bf B345} (1990) 301.}\
that the angles $\alpha$, $\beta$ and $\gamma$ as deduced {}from
the CP asymmetries in $B\ra\pi\pi$, $B\ra\psi K_S$ and $B_s\ra
\rho K_S$, respectively, would sum up to $\pi$ even if there is
new physics in $B^0$ mixing (this could be precisely the SUSY
contributions discussed in this subsection!)  such that the deduced
values of $\alpha$ and $\beta$ do not really correspond to angles of the
unitarity triangle.

Summarizing the phenomenological tests of the quark -- squark alignment
mechanism:
\item{(i)} Squarks are not necessarily degenerate;
\item{(ii)} $D-\bar D$ mixing is close to the experimental bound;
\item{(iii)} CP asymmetries in $B^0$ (but not $B_s$) decays may
differ by up to $\CO(0.3)$ {}from their Standard Model values.

\subsec{Higher Order Terms}

So far, we have considered only the dimension three mass terms that
arise when we assume a low energy effective model with an explicitly
broken horizontal symmetry, such that terms that break the symmetry by
$n\geq0$ ($n<0$) units of charge are suppressed by $\l^n$ (are
forbidden). With the same minimal set of assumptions, the effective
Lagrangian at low energies would contain also higher order terms that
obey similar selection rules. These terms may have important effects.
In particular, they may induce FCNC and affect quark -- squark
alignment. We now discuss these effects.

First, we consider constraints on four quark operators {}from FCNC.
Take, for example, a $\Delta s=2$ four quark operator $\CO_{K}$.
It would appear in the effective Lagrangian in the generic form
\eqn\genericff{{F_K\over M^2}\CO_K,}
where $F_K$ is a dimensionless coefficient that includes all suppression
factors, such as powers of $\l^n$ or powers of $1/(4\pi)^\ell$ if it
first appears at the $\ell$-loop level in the full theory, and $M$ is
the scale below which the effective Lagrangian description holds. Let us
further define
\eqn\defxk{X_K\equiv{\vev{K^0|\CO_K|\bar K^0}\over
\vev{K^0|(\bar s_L\gamma_\mu d_L)^2|\bar K^0}}.}
We similarly define the coefficients and matrix elements for $\Delta
c=2$ and $\Delta b=2$ operators. Then the requirement that four quark
operators do not contribute more than the experimental values of (or
bounds on) neutral meson mixing yields
\eqn\kdbff{M\geq\cases{\sqrt{X_KF_K}\ 1600\ TeV&$\Delta m_K$,\cr
\sqrt{X_K{\rm Im}(F_K)}\ 20000\ TeV&$\epsilon_K$,\cr
\sqrt{X_DF_D}\ 570\ TeV&$\Delta m_D$,\cr
\sqrt{X_BF_B}\ 530\ TeV&$\Delta m_B$.\cr}}

Since our framework incorporates Abelian horizontal symmetry, there is
no symmetry reason to forbid four quark operators of {\it e.g.} the form
$(Q_i\gamma^\mu Q_i^\dagger)(Q_j\gamma_\mu Q_j^\dagger)$, $i,j=1,2,3$
(and similarly for $\bar d_i$ and $\bar u_i$).  These terms are neutral
under $\CH$, and therefore are not suppressed by powers of $\l$.  When
rotating to the mass eigenbasis, FCNC operators are induced.  In
particular, some combination of $\Delta s=2$ and $\Delta c=2$ operators
is unavoidable. The weakest constraint corresponds to the case
$(V_L^d)_{12}=0$, $(V_L^u)_{12}\sim\l$ (as in models of quark -- squark
alignment).  It comes {}from $\Delta m_D$ with $X_D=1$ and $F_D=\l^2$
(assuming that the operator arises at tree level in the full theory):
\eqn\mffd{M\gsim\ 100\ TeV.}
If $(V_L^d)_{12}\sim\l$, as is the case in many of our models,
then a stronger bound {}from $\Delta m_K$ holds:
\eqn\mffk{M\gsim\ 300\ TeV.}
If, in addition, Im($F_K$) is comparable to the real part,
then an even stronger bound {}from $\epsilon_K$ holds:
\eqn\mffe{M\gsim\ 4000\ TeV.}

Additional and potentially stronger bounds arise if non-diagonal
four quark operators are not horizontally suppressed. For example,
there are four operators that may contribute to $K-\bar K$ mixing
(we take into account only $SU(2)_L\times U(1)_Y$ invariant terms):
\eqn\nondiagff{\eqalign{
\CO_{1K}=Q_2\gamma^\mu Q_1^\dagger Q_2 \gamma^\mu Q_1^\dagger,
&\ \ \ (X^{V+A}_K=1),\cr
\CO_{2K}=\bar d_2^\dagger\ \gamma^\mu \bar d_1\
\bar d_2^\dagger\ \gamma^\mu\bar d_1,
&\ \ \ (X^{V+A}_K=1),\cr
\CO_{3K}=Q_2 \gamma^\mu Q_1^\dagger\ \bar d_2^\dagger\gamma^\mu \bar
d_1,
&\ \ \ (X^{V-A}_K\approx3.6),\cr
\CO_{4K}=Q_2\ \bar d_1\ \ \bar d_2^\dagger\  Q_1^\dagger,
&\ \ \ (X^{S-P}_K\approx4.4).\cr}}
However, only $\CO_{3K}$ and $\CO_{4K}$ could avoid horizontal
suppression. This would happen if
\eqn\condndff{H_i (Q_1)-H_i (Q_2)=H_i (\bar d_1)-H_i (\bar d_2).}
$\CO_{1K}$ and $\CO_{2K}$ are always horizontally suppressed and
therefore give bounds that are not stronger than \mffk. Also, the
analogous $\Delta c=2$ and $\Delta b=2$ operators are always
horizontally suppressed. (This is simple to see, as a necessary result
of \condndff\ is $m_i/m_j\sim V_{ij}^2$.)  In models where the
operators $\CO_{3K}$ and $\CO_{4K}$ are not suppressed, the
following bounds apply (again, assuming that they are induced by tree
diagrams in the full theory):
\eqn\mffkb{M\gsim\ 3000\ TeV,}
and with CP violating phases of order one,
\eqn\mffeb{M\gsim\ 40000\ TeV.}

For the various explicit models presented in this section, we find

\item{(i)} In the master model, \condndff\ is fulfilled (see
\mastercharge) so that $M\gsim3000\ TeV$.

\item{(ii)} In the model of subsection 2.2, \condndff\ is not fulfilled
(see \modelacharge) so that $M\gsim300\ TeV$. However,
with a different choice of charge, $\bar d_1(0,1)$,
as made in \lns, \condndff\ is fulfilled and $M\gsim3000\ TeV$.

\item{(iii)} In the quark -- squark alignment model of subsection 2.3,
\condndff\ is not fulfilled (see \qsacharge) so that $M\gsim100\ TeV$.

Next we discuss bounds on the scale $M$ that arise {}from terms
involving higher powers of the scalar fields:
\eqn\xbreak{Q_i\bar d_j\phi_d\left({\phi_u\phi_d\over M^2}\right)^n;
\ \ \  Q_i\bar u_j\phi_u\left({\phi_u\phi_d\over M^2}\right)^m.}
These terms violate Natural Flavor Conservation and would
contribute to FCNC. In particular, for $i,j=1,2$, there would
be scalar-mediated tree diagrams contributing to $K-\bar K$ and
$D-\bar D$ mixing and therefore leading to bounds on $M$.

Note that the terms in \xbreak\ break $U(1)_X$ and therefore one cannot
use the simplified horizontal charge assignments achieved by
$U(1)_X\times U(1)_Y\times U(1)_B$ transformations. The allowed powers
$n$ and $m$ in \xbreak\ will depend on the true horizontal
symmetry. We consider here the case $n,m=1$. This leads to the
strongest possible bounds on the scale $M$ but in many of our models
$n,m>1$ and the bounds are consequently weaker.

The contributions depend on the intermediate scalar mass.
We take the upper bound on the mass of the lightest
neutral scalar in SUSY, $m_\phi\lsim150\ GeV$
\ref\haber{H.E. Haber and R. Hempfling, \PRL {\bf 66} (1991) 1815;
Y. Okada, M. Yamaguchi and T. Yanagida, Prog. Theo. Phys. {\bf 85}
 (1991) 1;
J. Ellis, G. Ridolfi and F. Zwirner, \PL {\bf B257} (1991) 83.}.
For $\tan\beta\sim1$, the strongest bound comes {}from $\Delta m_K$
(with $F_K\sim{9\over64}{v^4\over M^4}$), while for $\tan\beta\sim
m_t/m_b$, the strongest bound comes {}from $\Delta m_D$
(with $F_D\sim{9\over8}{v^4\over M^4}{m_b^2\over m_t^2}$):
\eqn\mxbreak{M\gsim\cases{26\ TeV&$\tan\beta\sim1$,\cr
4\ TeV&$\tan\beta\sim m_t/m_b$.\cr}}

The contributions to $M^d_{12}$, $M^d_{21}$ {}from terms of the form
\xbreak\ may spoil the precise alignment in our models
of quark -- squark alignment (QSA). Requiring that this should not
happen gives, in this class of models and if $n=1$ is allowed,
\eqn\xbreakqsa{M({\rm QSA})\gsim40\ TeV.}

When the bounds {}from four quark operators are taken into account, we
see that the effects of the $U(1)_X$ breaking terms on our mass
matrices, on FCNC and on quark -- squark alignment are always
unimportant.

\newsec{Beyond the Naive Relations}

\subsec{$|V_{ub}/V_{cb}|\ll|V_{us}|$}

As mentioned above, it may turn out that $|V_{ub}/V_{cb}|$ is not of the
same order of magnitude as $|V_{us}|$, namely that the naive
model-independent prediction of our framework fails. We now show that,
under special circumstances this naive prediction can be modified while
all other order of magnitude relations are maintained. We will present a
model where
\eqn\newmix{|V_{us}|\sim\l,\ \ |V_{cb}|\sim\l^2,\ \ |V_{ub}|\sim\l^4.}
We will see that to produce \newmix, we need a specific discrete
symmetry.

To suppress $|V_{ub}|$ below $\l^3$, certain entries in the mass
matrices have to be suppressed relative to their naive values.  Let us
first consider the (unrealistic) case of a continuous horizontal
symmetry, for which each entry in the mass matrix can either get its
naive value or vanish. Using appendix A we find that the mass matrices
have to take the following form
\eqn\requirevub{
M^d\sim\vev{\phi_d}\pmatrix{
\l^6&\l^5&0\cr \l^5&\l^4&\l^4\cr \l^3&0&\l^2},\ \ \
M^u\sim\vev{\phi_u}\pmatrix{
\l^6&0&0\cr \l^5&\l^3&\l^2\cr \l^3&\l&1}.}
Some additional entries may vanish, but not all of them.
In particular, to produce $|V_{us}|\sim\l$, we need $M^d_{12}\sim\l$.

As long as the effects of the discrete symmetry are negligible,
the mass matrices \requirevub\ give $|V_{ub}|\sim\l^5$.
If additional entries are zero, the value of $|V_{ub}|$ may be
even further suppressed, but it will always be an odd power of $\l$.
To obtain \newmix, the effects of the discrete symmetry have to
play a role. In particular, it must allow at least one of the
following four options:

(i) $M^u_{13}\sim\l^4$;
(ii) $M^u_{12}\sim\l^5$;
(iii) $M^d_{13}\sim\l^6$;
(iv) $M^d_{32}\sim\l^3$.

It is impossible to produce the required structure within a model with
a single $U(1)$. Considering models with $\CH=U(1)\times U(1)$,
we find that the required structure  cannot be produced in models
with $\ea\sim\l^2$ and $\eb\sim\l^3$, but it can in models of with
$\ea\sim\l$ and $\eb\sim\l^2$.
There is a very large number of models with this pattern of
symmetry breaking that produce \mixlam\ and \maslam. The charges of
all fields except $Q_3$ and $\bar u_3$ have more than one option.
For example, the charge of $\bar d_3$ could be either $(2,0)$
or $(0,1)$. Quite a few of these models give mass matrices of the
form \requirevub. We choose to present one example, to demonstrate
that the desired suppression of $|V_{ub}|$ is possible.

The model that we choose as an example has the following set of charges:
\eqn\modelbcharge{\matrix{Q_1&Q_2&Q_3&&\bar d_1&\bar d_2&\bar d_3&&
\bar u_1&\bar u_2&\bar u_3\cr
(-5,4)&(-2,2)&(0,0)&&(7,-2)&(6,-2)&(2,0)&&(11,-4)&(3,-1)&(0,0)\cr}.}
Actually, for $Q_1(-5,4)$ there is only a single choice for the charges
of all other fields except for $\bar d_1$ and $\bar u_1$.
The latter ones do not affect the required zeros. The choice
of their charges in \modelbcharge\ is correlated with the choice
of discrete symmetry below, and is motivated by considerations
that go beyond the low energy framework -- we explain this
in subsection 4.3.

We would like to introduce a discrete subgroup of the above symmetry
such that $|V_{ub}|\sim\l^4$. As mentioned above, one of the ways to do
it is to lift the zero in $M^u_{13}$ and have $M^u_{13}\sim\l^4$
instead. As the charge of $M^u_{13}$ under $U(1)_{H_1}\times U(1)_{H_2}$
is $(-5,4)$, there are two discrete subgroups that would do precisely
that: $Z_9\times Z_4$ and $Z_7\times Z_3$. With the latter symmetry (and
the choice of charges for $\bar d_1$ and $\bar u_1$ made in
\modelbcharge) we get the following mass matrices:
\eqn\modelbmas{
M^d\sim\vev{\phi_d}\pmatrix{
\ea^2\eb^2&\ea\eb^2&\ea^4\eb\cr \ea^5&\ea^4&\eb^2\cr
\eb&\ea^6\eb&\ea^2},\ \ \
M^u\sim\vev{\phi_u}\pmatrix{\ea^6&\ea^5&\ea^2\eb\cr
\ea^2\eb&\ea\eb&\ea^5\eb^2\cr \ea^4\eb^2&\ea^3\eb^2&1}.}
It is easy to check that these mass matrices produce the order of
magnitude relations \maslam\ and \newmix.

\subsec{Exact Relations between Quark Parameters}

The fact that an Abelian horizontal symmetry could produce zeros (or
highly suppressed terms) in the quark mass matrices, opens up the
interesting possibility of (close to) exact relations\foot{Clearly, with
a non-Abelian horizontal symmetry it is also possible to find exact
relations \philippe.} between various, otherwise independent, parameters
of the quark sector.

The model in the previous subsection is an example.
In general, $|V_{td}|$, $|V_{us}|$ and $|V_{cb}|$ are independent
parameters. Unitarity of the CKM matrix requires (see \wolfenstein)
\eqn\tduscbub{V_{td}=V_{us}^*V_{cb}^*-V_{ub}^*.}
In the previous subsection we presented models where
$|V_{ub}|\sim\l |V_{us}V_{cb}|$. Then, the following relation arises:
\eqn\tdexact{|V_{td}|=|V_{us}V_{cb}|\(1+\CO(\l)\).}
This relation is, of course, consistent with present constraints. (This
is a rather trivial statement because the best upper bound on $|V_{td}|$
at present comes {}from CKM unitarity.)  It is actually the only
phenomenologically acceptable relation that involves only mixing
parameters.

We searched for close-to-exact relations that involve both mass ratios
and mixing angles. Our basic assumption is that each entry has either
its ``naive" value as in eq. \mastermass, or it vanishes. (In case that
a discrete symmetry replaces a zero entry with one that is suppressed
compared to the naive one, the same relation would hold but potentially
with lesser accuracy.)  One can find some general rules. For example, no
exact relation can involve masses of first quark generation. The proof
for that is very simple: $m_d$ and $m_u$ depend on elements of the first
column in $M^d$ and $M^u$, respectively, but none of the mixing angles
depends on these elements to leading order (see Appendix A for the
dependence of the mixing angles on mass matrix elements).  Similar
considerations lead to the following conclusion:

Only a single exact relation could arise in our framework of
supersymmetric Abelian horizontal symmetry. It requires six entries in
the quark mass matrices to vanish:
\eqn\exactmas{M^d\sim\vev{\phi_d}\l^2\pmatrix{\l^6&0&\l^3\cr
\l^5&\l^2&0\cr \l^3&1&1\cr},\ \ \
M^u\sim\vev{\phi_u}\pmatrix{\l^6&0&0\cr \l^5&\l^3&0\cr \l^3&0&1\cr}.}
It is possible to exchange the second and third columns of $M^d$ without
changing the results. Note also that elements of the first columns do
not affect the relation, so some of the entries there might vanish as
well.

The exact relation that follows (to order $\l^2$) {}from \exactmas\ is
\eqn\exactrelation{{m_s^2\over m_b^2}=\left|{V_{cb}V_{ub}\over
V_{us}}\right|.}
Using the values  of mixing angles {}from \pdg\
and mass ratios {}from
\ref\gale{J. Gasser and H. Leutwyler, Phys. Rep. {\bf 87} (1992) 77.},
we have
\eqn\exactnumbers{
{m_s^2\over m_b^2}=0.0010^{+0.0009}_{-0.0006},\ \ \ \ \
\left|{V_{cb}V_{ub}\over V_{us}}\right|=0.0007^{+0.0006}_{-0.0004},}
so that the present accuracy in determining the various parameters does
not allow a test of the idea that an Abelian horizontal symmetry might
lead to exact relations.

\newsec{Spontaneously Broken $\CH$}

\subsec{Extending the Scalar Sector}

The low energy models described in the previous section can arise
naturally if $\CH$ is an exact symmetry of the Lagrangian, and is broken
by the VEV of a scalar that is a singlet of the Standard Model gauge
group and carries one unit of horizontal charge \frni.  For example, the
master model requires the existence of a single complex scalar field
$S(-1)$, with
\eqn\svev{\l={\vev{S}\over M}.}
($M$ is the scale at which the information about the spontaneous
symmetry breaking is communicated to the light fermions.)  The selection
rule becomes obvious now. For
\eqn\sumh{\eqalign{H(Q_i)+H(\bar d_j)=&l\geq0,\cr
                   H(Q_i)+H(\bar u_j)=&k\geq0,\cr}}
the exact horizontal symmetry allows only Yukawa terms of the form
\eqn\yuks{\CL_Y={\Gamma^d_{ij}\over M^l} S^l \phi_d Q_i\bar d_j
+{\Gamma^u_{ij}\over M^k} S^k \phi_u Q_i\bar u_j,}
(where the dimensionless Yukawa couplings $\Gamma^q_{ij}=\CO(1)$).
Supersymmetry requires that the Yukawa terms are analytic in $S$.
Consequently, $S^\dagger$ cannot take part in the Yukawa sector and
terms with $l<0$ or $k<0$ are forbidden (except for very highly
suppressed non-supersymmetric contributions).

Models with $\CH=U(1)_{H_1}\times U(1)_{H_2}$ need the introduction
of two Standard Model singlet scalars, $S_1$ and $S_2$, with horizontal
charges
\eqn\scharges{S_1(-1,0),\ \ \ \ \ S_2(0,-1),}
and vacuum expectation values
\eqn\svevs{{\vev{S_1}\over M}\sim\l^p,\ \ \ \ \
{\vev{S_2}\over M}\sim\l^q,}
where $(p,q)=(2,3)$ in the models of subsection 2.2 or $(p,q)=(1,2)$ in
the quark -- squark alignment models of subsection 2.3.  We always
assume that two separate scales of VEVs should break two different
symmetries, namely that all VEVs that break the same symmetry should be
at a single scale. Equations \scharges\ and \svevs\ are consistent with
this assumption, as each VEV breaks a different $U(1)$ factor in $\CH$.

\subsec{Bounds {}from FCNC}

The $S_i$ scalars couple non-diagonally to quarks. Consequently, they
mediate FCNC through tree diagrams. Specifically, they induce four quark
operators of the type $\CO_4$ of equation \nondiagff. As the masses of
the scalars is of the order of their vacuum expectation values, the
scale $M$ of equation \genericff\ should be replaced by $\vev{S}$.  The
factor $F$ is model dependent.  Let us examine a few examples.

\item{(i)} In the master model (see \mastermass),
$F_K\sim{m_dm_s\over\vev{S}^2}$, $F_B\sim{m_dm_b\over\vev{S}^2}$ and
$F_D\sim{m_um_c\over\vev{S}^2}$. Then the strongest bounds come {}from
the $K$ system,
\eqn\sboundmaster{\eqalign{
\vev{S}\gsim0.4\ TeV\ \Longrightarrow\ M\gsim2\ TeV,\ \ &(\Delta
m_K);\cr
\vev{S}\gsim1.4\ TeV\ \Longrightarrow\ M\gsim7\ TeV,\ \
&(\epsilon_K).\cr
}}

\item{(ii)} In the model of subsection 2.2 (see \modelamass),
$F_K\sim{\l^2 m_dm_s\over\vev{S_1}^2}$,
$F_B\sim{\l^4 m_dm_b\over\vev{S_1}^2}$ and
$F_D\sim{\l^4 m_um_c\over\vev{S_2}^2}$.
Then the only bound on $\vev{S_i}$ which is above the electroweak
scale comes {}from $\epsilon_K$,
\eqn\sboundmodela{
\vev{S_1}\gsim0.6\ TeV\ \Longrightarrow\ M\gsim13\ TeV,\ \
(\epsilon_K).}

\item{(iii)} In the model of reference \lns, all elements of $M^d$
assume their naive values, so that $F_K$ and $F_B$ are similar to the
master model. On the other hand, $M^u_{12}$ is suppressed and
consequently so is $F_D$. The resulting bounds are then
\eqn\sboundlns{\eqalign{\vev{S_2}\gsim0.4\ TeV\
\Longrightarrow\ M\gsim50\ TeV,\ \ &(\Delta m_K);\cr
\vev{S_2}\gsim1.4\ TeV\
\Longrightarrow\ M\gsim170\ TeV,\ \ &(\epsilon_K).\cr
}}

\item{(iv)} In the quark -- squark alignment models of subsection 2.3,
both $F_K$ and $F_B$ are always highly suppressed (see \mdqsa).
$F_D$ depends on the charge assignments of the $\bar u_i$ fields,
but in all cases $F_D\lsim{m_u m_c\over\vev{S_2}^2}$. If the bound
is saturated, then $\vev{S_2}$ cannot be lower than the electroweak
scale. If $F_D$ is further suppressed (as is the case in the example
given in eq. \ucharge), then no useful bound arises.

\subsec{QCD Anomalies and the Breaking of $U(1)_X$}

Now, that we have extended our framework to exact horizontal symmetries
that are only spontaneously broken, we should discuss in more detail the
subject of QCD anomalies\foot{We do not discuss $SU(2) \times U(1)$
anomalies because they depend on the charges in the lepton sector.
Since the mixing angles in that sector are not known, these charges are
not constrained significantly and anomalies can be easily avoided.}.  As
mentioned in section 2, QCD anomalies pose no problem because of the
$U(1)_X$ symmetry of the Yukawa sector. Furthermore, we mentioned that
$U(1)_X$ must be broken in some sector in the Lagrangian or else an
axion will be generated. We now discuss this in more detail.

We first discuss the case where $\CH\subset U(1)$ with the simplified
charges one gets by $U(1)_X\times U(1)_Y\times U(1)_B$ transformations.
This simplified horizontal symmetry may have a nonvanishing QCD anomaly
$A_H$:
\eqn\anomaly{A_H=\sum_i\(H(Q_i)+H(\bar u_i)+H(\bar d_i)\).}
The true horizontal symmetry is a $Z_n\subset U(1)_{\tilde H}\subset
U(1)_H\times U(1)_X$ and it must not have QCD anomaly,
\eqn\zerotanomaly{A_{\tilde H}=0({\rm mod}\ n).}
With $\tilde H$ charges given by
\eqn\tH{\tilde H=aH+bX,}
it is easy to see that $\tilde H$ and $H$ constrain the quark mass
matrices in precisely the same way. However, now
\eqn\tanomaly{A_{\tilde H}=aA_H+3b.}
We can always find appropriate values for $a$ and $b$ so that
\zerotanomaly\ is satisfied.

Having shown that $U(1)_X$ can always be used to avoid QCD anomalies,
we now discuss its breaking. We will assume that $U(1)_X$ is broken
by terms of the form
\eqn\xbroken{(\phi_d\phi_u)^p S^q.}
The $U(1)_H\times U(1)_X$ assignments of the scalar fields are
$\phi_d(0,-1)$, $\phi_u(0,0)$ and $S(-1,0)$. The interaction \xbroken\
should conserve only $Z_n\subset U(1)_{\tilde H}$. We therefore require
\eqn\tHxbreak{\tilde H((\phi_d\phi_u)^p S^q)=-aq-bp=0({\rm mod}\ n).}
Clearly, there are always  solutions $(p,q)$ to the requirement
\tHxbreak\ (take, for example $q=A_H({\rm mod}\ n)$ and $p=3$).

To summarize, QCD anomalies do not pose a problem in our framework:
even if $U(1)_H$ is anomalous, there is always a $Z_n\subset
U(1)_{\tilde H} \subset U(1)_H\times U(1)_X$ which is free of
QCD anomaly and should be considered as ``the true horizontal
symmetry.'' Note that the anomaly constraint restricts the $U(1)_X$
breaking terms, as these should be $Z_n$ invariant.

The extension of this mechanism to models where $\CH\subset U(1)\times
U(1)$ is straightforward. The horizontal symmetry is an anomaly free
$Z_m\times Z_n$ and the $U(1)_X$ symmetry is broken by a
$(\phi_d\phi_u)^p S_1^q S_2^r$ term. For example, consider the
model presented in subsection 3.1. This model was constructed to give
$|V_{ub}|\sim\l^4$. The charge assignments under the $U(1)_{H_1}\times
U(1)_{H_2}\times U(1)_X$ of the Yukawa sector are
\eqn\modelbcharge{\matrix{Q_1&Q_2&Q_3&&\bar d_1&\bar d_2&\bar d_3\cr
(-5,4,0)&(-2,2,0)&(0,0,0)&&(7,-2,1)&(6,-2,1)&(2,0,1)\cr
&&&&\bar u_1&\bar u_2&\bar u_3\cr
&&&&(11,-4,0)&(3,-1,0)&(0,0,0)\cr}.}
 We add a term
\eqn\xbreakb{(\phi_d\phi_u)S_1^5S_2}
which breaks $U(1)_{H_1}\times U(1)_{H_2}\times U(1)_X$ to
$Z_7\times Z_3$. Under this symmetry quarks transform as
\eqn\bcharge{\matrix{Q_1&Q_2&Q_3&&\bar d_1&\bar d_2&\bar d_3&&
\bar u_1&\bar u_2&\bar u_3\cr (2,1)&(5,2)&(0,0)&
&(2,0)&(1,0)&(4,2)&&(4,2)&(3,2)&(0,0)\cr},}
and the scalar fields as
\eqn\bscharge{\matrix{\phi_d&\phi_u&&S_1&S_2\cr
(-2,-2)&(0,0)&&(-1,0)&(0,-1)\cr}.}
The resulting mass matrices are those of equations \modelbmas.
It is straightforward to verify that the discrete symmetries are
free of QCD anomalies.

Note that in this model we could not choose an arbitrary discrete
symmetry -- the horizontal symmetry must be $Z_7\times Z_3$
to give the desired value for $|V_{ub}|$.  However, we used
the freedom in choosing the horizontal charges of $\bar d_1$ and
$\bar u_1$ and the choice of $p,q,r$ in equation \xbreakb, to find
solutions to the anomaly equations.

\newsec{Physics at the Scale $M$}

\subsec{Extending the Quark Sector}

In previous sections we have described the scale $M$ as the scale at
which the information about the breaking of the horizontal symmetry
$\CH$ is communicated to the light quarks, but we have not given any
explicit mechanism that would do that. In this section we make yet
another layer of assumptions: we use the mechanism suggested by
Froggatt and Nielsen (FN) \frni.

The FN mechanism assumes that there are additional quarks that transform
non-trivially under $\CH$.  These extra quarks come in mirror
representations, namely they may appear in any of the following
representations of $SU(3)_C\times SU(2)_L\times U(1)_Y\times U(1)_H$:
\eqn\mirror{\eqalign{
P(3,2,+1/6,H)\ \ {\rm and}&\ \ \bar P(\bar 3,2,-1/6,-H);\cr
D(3,1,-1/3,H)\ \ {\rm and}&\ \ \bar D(\bar 3,1,+1/3,-H);\cr
U(3,1,+2/3,H)\ \ {\rm and}&\ \ \bar U(\bar 3,1,-2/3,-H).\cr}}
Obviously, the new quarks can acquire heavy masses at a scale $M$ that
is much higher than the electroweak breaking scale.

At the scale $M$ we consider the most general $\CH$-invariant
renormalizable Yukawa terms.  As an example, we show how $M^u$ of
equation \modelamass,
\eqn\uamass{
M^u_{\rm light}\sim\vev{\phi_u}\pmatrix{\eb^2&0&\eb\cr \ea\eb&\eb&\ea\cr
\eb&0&1\cr},}
can be produced in the full theory \lns. We add to the light quarks
listed in eq. \modelacharge\ three $SU(2)$-singlet charge +2/3 mirror
quarks,
\eqn\mirroru{\matrix{U_1 & U_2 & U_3 \cr
(1,0) & (0,1) & (0,0) \cr},}
and $\bar U_i$ with opposite charges. Then the $6\times6$ matrix
$M^u_{\rm full}$ with columns corresponding to $(\bar u_i,\bar U_i)$ and
rows to $(Q_i,U_i)$ has order of magnitude entries
\eqn\mufull{M^u_{\rm full}=\pmatrix{
0 & 0 & 0 & 0 & \vev{\phi_u} & 0 \cr
0 & 0 & 0 & \vev{\phi_u} & 0 & 0 \cr
0 & 0 & \vev{\phi_u} & 0 & 0 & \vev{\phi_u} \cr
0 & \vev{S_2} & \vev{S_1} & M & 0 & \vev{S_1} \cr
0 & 0 & \vev{S_2} & 0 & M & \vev{S_2} \cr
\vev{S_2} & 0 & 0 &0 & 0 & M \cr}.}
When the heavy quarks at the scale $M$ are integrated out, the resulting
$M^u_{\rm light}$ for the three observed generations of up quarks is
$M^u$ of \uamass.

We had to add three $U+\bar U$ fields.  This could have been foreseen by
the following ``determinant argument'':
The determinant of the light fermions mass matrix is
\eqn\epdet{\det M^u_{\rm light}\sim\vev{\phi_u}^3\eb^3,}
which now means
\eqn\adetlight{\det M^u_{\rm light}\sim{\vev{\phi_u}^3\vev{S_2}^3
\over M^3}.}
However, an examination of the structure of $M^u_{\rm full}$ shows that
$\det M^u_{\rm full}$ is a polynomial in $\vev{\phi_u}$, $\vev{S_i}$
and $M$. As $\det M^u_{\rm full}=\det M^u_{\rm light}\times\det
M^u_{\rm heavy}$, we deduce that $\det M^u_{\rm heavy}\sim M^k$
with $k\geq3$, so that at least three $U+\bar U$ are required.

The general rule is then: if $\det M^q_{\rm light}\sim\vev{\phi_q}^3
\Pi_i\epsilon_i^{m_i}$, then the minimal number of massive $q$-quarks
required is $\sum_i m_i$. Thus, for example, as for the same model
(see \modelamass) $\det M^d_{\rm light}\sim\vev{\phi_d}^3\ea^6$,
at least six massive $D+\bar D$ (or $P+\bar P$) are required.
An explicit realization is given in ref. \lns. As another example,
in the master model $\det M^d_{\rm light}\sim\vev{\phi_d}\l^{12}$,
so at least twelve $D+\bar D$ are required, and
$\det M^u_{\rm light}\sim\vev{\phi_u}\l^{9}$,
so at least nine $U+\bar U$ are required.
The number of massive quarks will be important in our discussion of
Landau poles in subsection 6.1.

\subsec{Bounds on $M$ {}from FCNC}

With a full theory for physics at the scale $M$, we can check whether
the four quark operators discussed in subsection 2.4 indeed arise and
calculate the $F$ coefficients.  We find that, if baryon number is
conserved, the massive colored supermultiplets
cannot contribute to neutral meson mixing in tree diagrams.
Instead, the leading contributions come {}from box
diagrams with intermediate heavy $D$ or $U$ quarks and $S_i$ and
$\phi_q$ scalars.  An explicit calculation gives that the $F$
coefficients are suppressed by a factor $\sim1/4\pi$ compared to the
estimates in subsection 2.4.

The following bounds then hold on the scale $M$ of extra heavy quarks:

\item{(i)} Cabibbo mixing shows that the weakest bound that applies to
all models (coming {}from $\Delta m_D$) is
\eqn\fnbounda{M\gsim10\ TeV.}

\item{(ii)} In models where $(V_L^d)_{12}\sim\l$, a stronger bound
(coming {}from $\Delta m_K$) holds,
\eqn\fnboundb{M\gsim25\ TeV.}

\item{(iii)} In models where Im$\((V_L^d)_{11}(V_L^d)^*_{12}\)\sim\l$,
an even stronger bound (coming {}from $\epsilon_K$) holds,
\eqn\fnboundc{M\gsim300\ TeV.}

\item{(iv)} In models where $H(Q_1)-H(Q_2)=H(\bar d_1)-H(\bar d_2)$,
a bound {}from $\Delta m_K$ stronger than \fnboundb\ holds,
\eqn\fnboundd{M\gsim\ 250\ TeV.}

\item{(v)} In models where $H(Q_1)-H(Q_2)=H(\bar d_1)-H(\bar d_2)$,
and there are CP violating phases of order one, a bound {}from
$\epsilon_K$ stronger than \fnboundc\ holds,
\eqn\fnbounde{M\gsim\ 3000\ TeV.}

This can be easily applied to the explicit models of section 2.  The
master model is constrained by \fnboundd\ and possibly \fnbounde.  The
model of subsection 2.2 is constrained by \fnboundb\ and possibly
\fnboundc\ (but its version presented in \lns\ has the same constraints
as the master model). Models of quark -- squark alignment are
constrained by \fnbounda.

\newsec{Physics Above $M$}

\subsec{Landau Poles}

The explanation of the physics responsible for the hierarchy in the
quark sector parameters is now complete. It involves two scales (beyond
the electroweak breaking scale): the scale of spontaneous
$\CH$-breaking, $\vev{S}$ (this might happen in several scales), and the
higher scale at which the information is communicated to the observed
quarks, $M$. Physics above the scale $M$ has no direct bearing on the
quark parameters. It may, however, further constrain the scale $M$.

These constraints on the scale $M$ are a result of the running of the
coupling constants: we do not allow Landau poles below the Planck scale
$M_P$. Landau poles may arise when we add massive supermultiplets that
transform non-trivially under $SU(3)_C\times SU(2)_L\times U(1)_Y$.  In
our full framework, as described in sections 2--5, we have, in addition
to the representations of the minimal supersymmetric Standard Model, the
extra heavy quarks required for the FN mechanism.  To calculate the
running of the coupling constants up to the Planck scale, we need to
know also the particle representations and the gauge structure between
$M$ and $M_P$. If we adopt the most conservative approach, namely that
the gauge group is $SU(3)_C\times SU(2)_L\times U(1)_Y$ up to $M_P$, we
get a lower bound on the scale $M$, that we denote by $M_{\rm min}$.
Alternatively, for a given scale $M$ we can get an upper bound on the
scale at which the gauge symmetry has to increase -- below the location
of the Landau pole. Below, we present this upper bound corresponding to
$M\sim250\ TeV$ and denote it by $M^G_L$.


In our framework, the one loop running of the
three gauge couplings (neglecting threshold effects) is given by
\eqn\alpharun{\eqalign{\(\alpha_s(M_P)\)^{-1}=&\(\alpha_s(m_Z)\)^{-1}
+{7\over2\pi}\ln{M_{\rm SUSY}\over m_Z}
+{3\over2\pi}\ln{M\over M_{\rm SUSY}}
+{3-N_3\over2\pi}\ln{M_P\over M},\cr
\(\alpha_2(M_P)\)^{-1}=&\(\alpha_2(m_Z)\)^{-1}
+{3\over2\pi}\ln{M_{\rm SUSY}\over m_Z}
-{1\over2\pi}\ln{M\over M_{\rm SUSY}}
-{1+N_2\over2\pi}\ln{M_P\over M},\cr
\(\alpha_1(M_P)\)^{-1}=&\(\alpha_1(m_Z)\)^{-1}
-{41\over20\pi}\ln{M_{\rm SUSY}\over m_Z}
-{33\over10\pi}\ln{M\over M_{\rm SUSY}}
-{33+N_1\over10\pi}\ln{M_P\over M},\cr}}
where
\eqn\defncly{\eqalign{N_3=&\ 2N_P+N_U+N_D,\cr
N_2=&\ 3N_P+N_L,\cr N_1=&\ N_P+8N_U+2N_D+3N_L+6N_E,\cr}}
with $N_P$ the
number of mirror quark doublets, $N_U$ the number of mirror up-quark
singlets, $N_D$ the number of mirror down-quark singlets, $N_L$ the
number of mirror lepton doublets and $N_E$ the number of charged lepton
singlets.  For the gauge couplings at the scale $m_Z$, we take
\eqn\alphamz{\(\alpha_s(m_Z)\)^{-1}\approx 9,\ \ \
\(\alpha_2(m_Z)\)^{-1}\approx 30,\ \ \
\(\alpha_1(m_Z)\)^{-1}\approx 59.}
(Note that $\(\alpha_1(m_Z)\)^{-1}={3\over5}\(\alpha(m_Z)\)^{-1}
\cos^2\theta_W$ is defined differently {}from $\alpha^\prime$.)

The requirement that there is no Landau pole below $M_P$ gives the
following bounds:

\vskip 1cm
\begintable
\multispan{3}\tstrut\hfil 1. No Landau Poles in $\alpha_s$ \hfil\crthick
 $N_3$  | $M_{\rm min}\(TeV\)$ | $M_L^{SU(3)}\(TeV\)$ \cr
 $5$  | $1$                  |                              \cr
 $6$  | $5\cdot10^2$         |    $2\cdot10^{15}$           \cr
 $7$  | $4\cdot10^4$         | $1\cdot10^{12}$              \cr
 $8$  | $1\cdot10^6$         | $1\cdot10^{10}$              \cr
 $9$  | $1\cdot10^7$         | $7\cdot10^{8}$ \endtable

\vskip 1cm
\begintable
\multispan{3}\tstrut\hfil 2. No Landau Poles in $\alpha_2$ \hfil\crthick
 $N_2$ | $M_{\rm min}\(TeV\)$ | $M_L^{SU(2)}\(TeV\)$ \cr
 $5$   | $3\cdot10^2$ | $1\cdot10^{16}$ \cr
 $6$   | $5\cdot10^4$ | $1\cdot10^{14}$ \cr
 $7$   | $2\cdot10^6$ | $5\cdot10^{12}$ \cr
 $8$   | $3\cdot10^7$ | $4\cdot10^{11}$ \cr
 $9$   | $3\cdot10^8$ | $4\cdot10^{10}$ \endtable

\vskip 1cm
\begintable
\multispan{3}\tstrut\hfil 3. No Landau Poles in $\alpha_1$ \hfil\crthick
 $N_1\quad$ | $M_{\rm min}\(TeV\)$ | $M_L^{U(1)}\(TeV\)$ \cr
 $15$  | $0.3$        | \cr
 $16$  | $3         $ | \cr
 $17$  | $26        $ | \cr
 $18$  | $2\cdot10^2$ | \cr
 $19$  | $9\cdot10^2$ | $8\cdot10^{15}$ \cr
 $20$  | $4\cdot10^3$ | $4\cdot10^{15}$ \cr
 $25$  | $1\cdot10^6$ | $3\cdot10^{14}$ \cr
 $30$  | $6\cdot10^7$ | $3\cdot10^{13}$ \endtable

\subsec{Could There Be Low Energy Flavor Physics?}

The reason that we have studied the various constraints on the scales so
carefully is that we consider the following question as highly
important: Could the flavor physics, responsible for the hierarchy in
the quark sector parameters, be directly observable?  Without going into
a detailed discussion of the possible signatures, let us just estimate
for now that, in order that the New Physics is observed, at least the
lowest of the new scales (typically the scale below which the horizontal
symmetry is completely broken) should be at the few TeV region.
Explicitly, we ask whether we could have $\vev{S_2}\lsim2\ TeV$
($\vev{S}\lsim2\ TeV$ for a single $Z_n$). As $\vev{S_2}/M\gsim\l^3$ in
our various models, we should check whether $M\lsim250\ TeV$ is allowed.

Examining the three Tables above we see that to have low energy flavor
physics (and no modified gauge symmetry below $M_P$), we need
$N_3\leq6$, $N_2\leq5$ and $N_1\leq18$. These constraints are very
difficult to satisfy. For example, we can only allow $N_P\leq1$ and
$N_U\leq2$. It is straightforward to see that in all our models, to
explain $m_u\ll m_c\ll m_t$, at least three massive up quarks are
required; to explain $m_d\ll m_s\ll m_b\ll m_t$ with $\tan\beta\sim1$ at
least five massive down quarks are required; and if we wish to explain
the hierarchy in lepton masses in a similar way, a large number of
massive charged leptons (at least six with $\tan\beta\sim1$) is required
as well. Then it is hopeless, under our assumptions, to have low energy
flavor physics.

One way out of this grim conclusion is to give up the most speculative
part in our theory, for example, allow a change in the gauge structure
not too far above $M$. Then, we should not worry about Landau poles.
The constraints {}from FCNC are much weaker and allow many of our models
to have flavor physics at the TeV scale. For example, in all models of
quark -- squark alignment, if we ignore the Landau poles constraints,
$M$ of order 10 $TeV$ and $\vev{S}$ below $TeV$ are allowed.

But for now, let us adopt our full framework, namely spontaneously
broken $\CH$ at a scale $\vev{S_i}$, mirror quarks at $M$, and neither
new particles nor new gauge structure above $M$ and up to $M_P$. Then,
to allow low energy flavor physics, we have to give up some of the
ingredients in our models that necessitated the large number of massive
supermultiplets. First, we should better work with $\CH=Z_m\times Z_n$
symmetry and $\ea\sim\l^2$, $\eb\sim\l^3$ breaking parameters. This, as
explained in subsection 5.1, allows lower powers of $\epsilon_i$ in the
determinant and hence fewer massive quarks. However, in addition we have
to modify our estimates of two parameters:

\item{(i)} $m_u=0$. \hfill\break
What we mean here is not that the bare mass of the up quark is highly
suppressed but finite - that would require many more massive $U$s. We
need a bare mass that is {\it exactly} zero (up to non-perturbative QCD
effects that would generate its value as determined {}from meson
masses). Then, the FN mechanism should account for $m_c\ll m_t$ only,
which can be done with a single massive $U$ or $P$.

\item{(ii)} $\tan\beta\sim m_t/m_b$. \hfill\break For a large
$\tan\beta$, the FN mechanism is not responsible for $m_b/m_t$. To
account for $m_d\ll m_s\ll m_b$, only three massive $D$s or $P$s are
needed.  (For recent discussions of how to naturally produce
$\tan\beta\gg1$, see
\ref\largetb{L.J. Hall, R. Rattazzi and U. Sarid, LBL preprint
LBL-33997 (1993), {\rm hep-ph}-9306309; A. Nelson and L. Randall,
San Diego preprint UCSD-PTH-93-24 (1993), {\rm hep-ph}-9308277.}.)

Let us give an explicit example \lns. Take the model presented in
subsection 2.2 and modify it to the case $m_u=0$, $\tan\beta\sim
m_t/m_b$. This can be done, for example, by modifying the charges in
\modelacharge\ to
\eqn\leacharge{\matrix{Q_1&Q_2&Q_3&&\bar d_1&\bar d_2&\bar d_3&&
\bar u_1&\bar u_2&\bar u_3\cr
(0,1)&(1,0)&(0,0)&&(2,-1)&(0,0)&(0,0)&&(-1,1)&(-1,1)&(0,0)\cr}}
The mass matrices have the order of magnitude entries
\eqn\leamass{
M^d\sim\vev{\phi_d}\pmatrix{\ea^2&\eb&\eb\cr
0&\ea&\ea\cr 0&1&1\cr},\ \ \
M^u\sim\vev{\phi_u}\pmatrix{0&0&\eb\cr \eb&\eb&\ea\cr 0&0&1\cr}.}
Note that $M^u$ has a zero eigenvalue. Then
\eqn\leadet{\det M^d\sim\vev{\phi_d}^3\ea^3,\ \ \
m_c m_t\sim\vev{\phi_u}^2\eb,}
imply that a full theory could be constructed with a single $U$ and
three $D$s. We choose to construct the massive sector with one doublet
$P$ and three $D$s:
\eqn\heacharge{\matrix{P&&D_1&D_2&D_3\cr
(1,-1)&&(1,0)&(0,1)&(-1,1).\cr}}
The $4\times4$ matrix $M^u_{\rm full}$ with the fourth row (column)
corresponding to $P(\bar P)$ is
\eqn\lpmufull{M^u_{\rm light}\sim\pmatrix{0&0&0&0\cr 0&0&0&\vev{S_2}\cr
0&0&\vev{\phi_u}&0\cr 0&\vev{\phi_u}&0&M\cr}\ \Longrightarrow\
M^u_{\rm light}\sim\vev{\phi_u}\pmatrix{0&0&0\cr 0&\eb&0\cr 0&0&1\cr},}
while the $6\times6$ matrix $M^d_{\rm full}$ with rows corresponding
to $(Q_i,D_i)$ and columns to $(\bar d_i,\bar D_i)$ is
\eqn\lpmdfull{M^d_{\rm full}\sim\pmatrix{0&0&0&0&\vev{\phi_d}&0\cr
0&0&0&\vev{\phi_d}&0&0\cr 0&\vev{\phi_d}&\vev{\phi_d}&0&0&0\cr
0&\vev{S_1}&\vev{S_1}&M&0&0\cr 0&\vev{S_2}&\vev{S_2}&0&M&\vev{S_1}\cr
\vev{S_1}&0&0&0&0&M\cr},}
leading to $M^d_{\rm light}$ of eq. \leamass.

In this model,
\eqn\countM{N_3=5,\ \ N_2=3+N_L,\ \ N_1=7+3N_L+6N_E.}
The constraints {}from the Landau poles do not exclude $M$ in the few
hundreds TeV region, though with $N_L=2$ and $N_E=1$ the bound would
rise to 900 $TeV$. The Landau poles constraints are even weaker for the
model of \lns\ with $m_u=0$ and $\tan\beta\sim m_t/m_b$: there we could
use the FN mechanism with one massive doublet and two massive down
singlets, which would allow the addition of three massive charged
leptons with $M\sim300\ TeV$. However, in the latter case the bounds
{}from four Fermi operators are between 250 $TeV$ (from $\Delta m_K$)
and 3000 $TeV$ (from $\epsilon_K$), depending on the phase.

To summarize, models with $\CH\subset U(1)_{H_1}\times U(1)_{H_2}$ and
breaking parameters $\ea\sim\l^2$, $\eb\sim\l^3$, are viable candidates
for low energy flavor physics ($\vev{S_2}\sim2\ TeV$) provided that (i)
$m_u=0$ (at high energies) and (ii) $\tan\beta\sim m_t/m_b$. Scanning
all such models, we found only two examples where, under special
circumstances, $\vev{S_2}$ can be at the $TeV$ scale: the model
of subsection 2.2, if lepton masses do not all come {}from FN mechanism,
and the model of reference \lns, if CP violating phases in box diagrams
involving the massive quarks are suppressed.  Also, if the gauge
structure changes above $M$, then many of our models could, in
principle, have their flavor physics at low energy.

\newsec{Small Hierarchy {}From Large Hierarchy}

Let us assume that the horizontal symmetry is
\eqn\zsubn{Z_n\subset U(1)_H\times U(1)_X.}
If $U(1)_X$ is broken by a term $S^{n-2}(\phi_u\phi_d)/M_p^{n-3}$ in the
superpotential $W$, then all the following terms will also appear in $W$:
\eqn\superpot{W={1\over M_P^{n-3}}\sum_{m=0}^{\(n/2\)} A_m S^{n-2m}
(\phi_u\phi_d)^m,}
where $A_m$ are dimensionless coefficients of order one.  This
contributes
to the Higgs potential, $V^W=|\partial W/\partial S|^2 +|\partial
W/\partial\phi_u|^2+|\partial W/\partial\phi_d|^2$:
\eqn\wtov{
\eqalign{V^W={1\over M_P^{2n-6}}\sum_{m,k=0}^{\(n/2\)}&A_m A_k\left\{
(n-2m)(n-2k)S^{2(n-m-k-1)}(\phi_u\phi_d)^{m+k}\right.\cr +&\left.
mkS^{2(n-m-k)}(\phi_u\phi_d)^{m+k-2}(\phi_u^2+\phi_d^2)\right\}.\cr}}
In addition there are SUSY $D$-terms,
\eqn\dtov{
V^D=D_d|\phi_d|^4+D_u|\phi_u|^4+D_{ud}|\phi_u|^2|\phi_d|^2,}
soft SUSY breaking terms analytic in the fields,
\eqn\mutov{V^\tm={\tm\over M_P^{n-3}}\sum_{m=0}^{\(n/2\)}B_m S^{n-2m}
(\phi_u\phi_d)^m.}
and $A$-terms,
\eqn\atov{V^A=\tm^2(C_s|S|^2+C_u|\phi_u|^2+C_d|\phi_d|^2).}
$D_i$, $B_i$ and $C_i$ are all dimensionless coefficients of order one,
and $\tm$ is the SUSY breaking scale.

The extremum equations,
\eqn\extremum{{\partial V\over\partial S}=0,\ \
{\partial V\over\partial \phi_u}=0,\ \
{\partial V\over\partial \phi_d}=0.}
have a solution of the form:
\eqn\vevs{\vev{S}\sim\ M_P\left({\tm\over M_P}\right)^{1\over n-2},}
\eqn\vevphi{\vev{\phi_u}\ \sim\vev{\phi_d}\sim\tm.}

It is easy to check that, even though the potential for the various
scalars is very flat, all scalars acquire masses larger than or of order
$\tm$.  There might, however, be cosmological problems with such scalars
whose couplings are very weak.

This solution has some attractive features:

\item{(i)} The so-called $\mu$-problem in SUSY is solved. The horizontal
symmetry
forbids a term $\phi_u\phi_d$ in the superpotential. Equation \vevphi\
implies that the electroweak breaking scale is naturally at the SUSY
breaking scale.

\item{(ii)} Out of the large hierarchy between the SUSY breaking scale
$\tm$ and the Planck scale $M_P$, we can naturally produce a smaller
hierarchy of scales $\vev{S_i}$, as can be seen {}from \vevs.

As an example, consider the models of reference \lns, described in
subsection 2.2. There we need hierarchy of scales:
\eqn\scaleslns{\vev{S_2}\ :\ \vev{S_1}\ :\ M\ \sim\ 1\ :\ 5\ :\ 125.}
Let us assume that the scale $M$ is the spontaneous breaking scale
of some additional discrete symmetry by a vacuum expectation value of
a scalar field $\vev{S_3}$. Then, if the full discrete symmetry
is $Z_7\times Z_6\times Z_{10}$, we get {}from \vevs
\eqn\lnsvevs{\eqalign{
\vev{S_2}\sim&M_P\left({\tm\over M_P}\right)^{1/4}\sim10^{15}\ GeV,\cr
\vev{S_1}\sim&M_P\left({\tm\over M_P}\right)^{1/5}\sim6\times
10^{15}\ GeV,\cr
\vev{S_3}\sim&M_P\left({\tm\over M_P}\right)^{1/8}\sim10^{17}\ GeV,\cr}}
consistent with \scaleslns.

Note that if the relevant large scale is indeed $M_P$, as we assumed in
this subsection, it is very difficult to produce a low energy
horizontal symmetry. With $n=3$ we get $\vev{S}=\tm$, but for $n\geq4$
we get $\vev{S}\geq\sqrt{\tm M_P}\sim10^{8}\ TeV$.

It could in principle be that the scale in the denominator
of the non-renormalizable terms of the Higgs potential is
lower than $M_P$. It could even be that there is a ladder of
scales, one for each broken symmetry. This would of course
enable one to produce a hierarchy of relatively low scales.

\newsec{Conclusions}

Abelian horizontal symmetries could explain in a simple and natural way
the smallness of the quark sector parameters and the hierarchy among
them. The master model, presented in subsection 2.1, demonstrates that a
simple Abelian group, with a single not-so-small breaking parameter, and
``reasonable" horizontal charges, could account for the fact that the
hierarchy in the quark sector parameters spans five orders of magnitude.

It would be much more difficult, however, to have convincing evidence
that such a symmetry is indeed responsible for the hierarchy. As far as
the quark mass ratios and mixing angles are concerned, the symmetry
explains eight order of magnitude relations but predicts only one. The
fact that this single prediction is indeed consistent with present
measurements is encouraging but can hardly be taken as evidence for the
horizontal symmetry idea.

The simplest way in which the existence of a horizontal symmetry could
be revealed is by direct observation of new particles related to flavor
physics. Though not absolutely necessary, we find that it is likely that
the mechanism that produces the hierarchy in the quark parameters
requires the existence of extra scalars with flavor changing couplings
and massive mirror quarks. A crucial question is then whether these new
particles could have masses at scales accessible to experiment\foot{We
should note here that if $\CH$ is spontaneously broken at low energies,
there might be cosmological problems with domain walls.}, say a few
$TeV$.  We find that this is not a very likely possibility but not
impossible.  If we are fortunate to have flavor physics at low energies,
it probably means that the bare mass of the up quark vanishes (thus
providing a solution to the strong CP problem) and that $\tan\beta$ is
large.

Whatever scale we associate with the New Physics, it may have many other
interesting consequences:

\item{(i)} A horizontal symmetry could align quark mass matrices with
squark mass-squared matrices in a precise enough way to suppress SUSY
contributions to neutral meson mixing. If squarks are found, and if they
are non-degenerate, a horizontal symmetry is almost unavoidable.
Another crucial test to the quark -- squark alignment mechanism is that
$D-\bar D$ mixing should be close to the present experimental upper
bound.

\item{(ii)} An Abelian horizontal symmetry could lead to an exact
relation between the parameters, $m_s^2/m_b^2= |V_{cb} V_{ub}/V_{us}|$.

\item{(iii)} A horizontal discrete symmetry has a natural mechanism to
generate the hierarchy among scales of spontaneous symmetry breaking.
It can solve in a simple way the $\mu$-problem of supersymmetry and
provide a natural explanation to the fact that the electroweak breaking
scale is close to the SUSY breaking scale.

While none of these possibilities is a necessary consequence of a
horizontal symmetry, they may provide support to the idea that the
hierarchy in the quark sector parameters is a result of such a symmetry.

\bigbreak
\centerline{{\bf Acknowledgements}}

It is a pleasure to thank T. Banks, A. Dabholkar, M. Dine, K.
Intriligator, D. Kaplan, A. Nelson, P. Pouliot and S. Shenker for
several useful discussions.  This work was supported in part by DOE
grant \#DE-FG05-90ER40559. YN is an incumbent of the Ruth E.  Recu
Career Development chair, and is supported in part by the Israel
Commission for Basic Research, by the United States -- Israel Binational
Science Foundation (BSF), and by the Minerva Foundation.

\appendix{A}{Diagonalizing the Quark Mass Matrices}

We would like to estimate the elements of the diagonalizing matrices
$V_M^q$ ($M=L$ or $R$, $q=u$ or $d$) of the mass matrices as defined in
\vquark. We follow  the formalism of reference
\ref\hara{L.J. Hall and A. Rasin, \PL {\bf B315} (1993) 164.}
with some modifications which are appropriate for our case.  We present
the $V^q_L$ matrices as
\eqn\Vi{V^q_L=
\pmatrix{1 & -s^q_{12} & 0 \cr s^{q*}_{12} & 1 & 0 \cr 0 & 0 & 1\cr}
\pmatrix{1 & 0 & -s^q_{13} \cr 0 & 1 & 0 \cr s^{q*}_{13} & 0 & 1\cr}
\pmatrix{1 & 0 & 0 \cr 0 & 1 & -s^q_{23} \cr 0 & s^{q*}_{23} & 1\cr}.}
$V^u_R$ is given by a similar formula, with
$s^u_{jk}$ replaced by $\psu_{jk}$. $V^d_R$ is
\eqn\VdR{V^d_R=
\pmatrix{1 & -\psd_{12} & 0 \cr s^{\prime d*}_{12} & 1& 0 \cr 0&0&1\cr}
\pmatrix{1 & 0 & -\psd_{13} \cr 0&1&0\cr s^{\prime d*}_{13} & 0 & 1\cr}
\pmatrix{1&0&0\cr 0 & c^{\prime d}_{23} & -\psd_{23} \cr
         0 & s^{\prime d*}_{23}& c^{\prime d*}_{23}\cr}.}
The difference between $V^d_R$ and the other $V$'s stems {}from the fact
that we allow $M^d_{i2} \sim M^d_{i3}$ ($i=1,2,3$), as is the case in
many of our models. Below we give the leading contributions to the
diagonalizing parameters $s^i_{jk}$ and $s^{\prime i}_{jk}$ in terms of
the mass matrices.

Starting with the up sector, we assume that there is a hierarchy
between all rows and all columns and define:
\eqn\yudef{y^u_{ij}\equiv {M^u_{ij}\over M^u_{33}}.}
Following \hara, we also introduce the notation:
\eqn\tyudef{\tilde y^u_{22}=y^u_{22}y^u_{33}-y^u_{23}y^u_{32},}
($|\tilde y_{22}|=m_c/m_t$).
Then, the $s^u_{ij}$ mixing angles are
\eqn\suij{\eqalign{
s_{12}^u=&{y^u_{12}\over \tilde y^u_{22}}
 + {y^u_{11}y^{u*}_{21}\over|\tilde y^{u*}_{22}|^2}
 -{y^u_{13}(y^u_{32}+y^{u*}_{23}y^u_{22})\over \tilde y^u_{22}}
 -{y^u_{11}y^{u*}_{31}(y^{u*}_{23}+y^u_{32}y^{u*}_{22})
   \over|\tilde y^{u*}_{22}|^2},\cr
s_{13}^u=&y^u_{13} + y^u_{11}y^{u*}_{31} +
y^u_{12}(y^{u*}_{32}+y^{u*}_{22}y^u_{23}) +
y^u_{11}y^{u*}_{21}(y^u_{23}+y^{u}_{22}y^{u*}_{32}),\cr
s_{23}^u=&y^u_{23}+y^u_{22}y^{u*}_{32}.\cr}}
The $s^{\prime u}_{ij}$ mixing angles are given by formulae similar to
\suij, with the replacement $y^u_{ij}\leftrightarrow y^{u*}_{ji}$.

Turning to the down sector, we define
\eqn\ydij{\eqalign{
y_{i1}^d=&{M_{i1}^d\over{\sqrt{|M_{22}^d|^2+|M_{33}^d|^2}}},\cr
y_{i2}^d=&{M_{i2}^dM_{33}^d-M_{i3}^dM_{32}^d\over
          |M_{22}^d|^2+|M_{33}^d|^2}, \cr
y_{i3}^d=&{M_{i3}^dM_{33}^{d*}+M_{i2}^dM_{32}^{d*}\over
          |M_{22}^d|^2+|M_{33}^d|^2}. \cr}}
 Then, the $s^d_{ij}$ mixing angles are
\eqn\sdij{\eqalign{
s_{12}^d=&{y^d_{12}\over y^d_{22}}
 +{y^d_{11} y^{d*}_{21}\over|y^{d*}_{22}|^2} - y^d_{13}y^{d*}_{23}
 -{y^d_{11}y^{d*}_{31}y^{d*}_{23}\over|y^d_{22}|^2},\cr
s_{13}^d=&y^d_{13} + y^d_{11} y^{d*}_{31}
  + y^d_{12} y^{d*}_{22} y^d_{23} +y^d_{11}y^{d*}_{21}y^d_{23},\cr
s_{23}^d=&y^d_{23}.\cr}}
The $s^{\prime d}_{ij}$ mixing angles are
\eqn\spdij{\eqalign{
s_{12}^{\prime d}=&{y^{d*}_{21}\over  y^{d*}_{22}}
  + {y^{d*}_{11} y^d_{12}\over |y^d_{22}|^2}
  -{y^{d*}_{31}y^{d*}_{23}\over  y^{d*}_{22}}
  -{y^{d*}_{11} y^d_{13} y^{d*}_{23} \over  y^{d*}_{22}},\cr
s_{13}^{\prime d}=&y^{d*}_{31} + y^{d*}_{11}y^d_{13}
   + y^{d*}_{21}y^d_{23} + y^{d*}_{11}y^d_{12}y^{d*}_{22}y^d_{23},\cr
s_{23}^{\prime d}=&{M_{32}^{d*}\over \sqrt{|M_{32}^d|^2+|M_{33}^d|^2}}
   +y^{d*}_{22}y^d_{23},\cr
c_{23}^{\prime d}=&{M_{33}^{d*}\over \sqrt{|M_{32}^d|^2+|M_{33}^d|^2}}.
\cr}}

If $M_{32}=0$ (as happens in most of our models), one may
alternatively define: $y^d_{ij}=M^d_{ij}/M^d_{33}$. Then \sdij\ and
\spdij\ still hold, except the last equation in \spdij\ which should be
replaced by $c^{\prime d}_{23}=1$.
We stress that the above formulae give the {\it leading}
contribution to the mixing parameters (when expanding in powers of
$\l\sim0.2$), and should not be used to extract the next to leading
terms, as these may have further contributions.

The CKM matrix elements are then given, to leading order, by:
\eqn\vij{\eqalign{|V_{us}|=&|s_{12}^d-s_{12}^u|,\cr
  |V_{cb}|=&|s_{23}^d-s_{23}^u|,\cr
  |V_{ub}|=&|s_{13}^d-s_{13}^u-s_{12}^u(s_{23}^d-s_{23}^u)|.\cr}}

\appendix{B}{Subtlety in the kinetic terms}

So far we ignored the potential renormalization of the kinetic
terms.\foot{We thank A. Dabholkar for raising this subject.} The
canonical kinetic terms can be modified to
\eqn\kin{\sum_{q,i,j} R^q_{ij}q^\dagger_i \gamma^\mu \partial_\mu q_j }
($q=Q, \bar u, \bar d$ and $i=1,2,3$) in a way consistent with the
horizontal symmetry (we assume $\CH = U(1)_{H_1} \times U(1)_{H_2}$ with
two explicit breaking parameters $\ea$ and $\eb$):
\eqn\kcons{ R^q_{ij} \sim \ea^{|H_1(q_i) -
H_1(q_j)|}\eb^{|H_2(q_i) - H_2(q_j)|}. }
The mass terms in the Lagrangian, $M^u_{ij}$ and $M^d_{ij}$,
are constrained by the symmetries:
\eqn\mdcons{M^d_{ij} \sim \ea^{H_1(Q_i) + H_1(\bar
d_j)}\eb^{H_2(Q_i) + H_2(\bar d_j)}}
when both $H_1(Q_i) + H_1(\bar d_j) \ge 0$ and $H_2(Q_i) + H_2(\bar
d_j)\ge 0 $ and zero otherwise and
\eqn\mucons{M^u_{ij} \sim\ea^{H_1(Q_i) + H_1(\bar
u_j)}\eb^{H_2(Q_i) + H_2(\bar u_j)} }
when both $H_1(Q_i) + H_1(\bar u_j) \ge 0 $ and $H_2(Q_i) + H_2(\bar
u_j)\ge 0 $ and zero otherwise.

We can rescale the $q_i$'s and mix different $q_i$'s with the same
horizontal charges to set all diagonal elements of $R^q$ to one and all
off diagonal elements with $H_1(q_i) - H_1(q_j) = H_2(q_i) - H_2(q_j)=0$
to zero.  Then
\eqn\litr{R^q_{ij} = \delta_{ij} +r^q_{ij} }
where $r^q_{ij} = \CO(\ea, \eb)$.

In order to find the true mass matrices, the fields $q_i$ should be
redefined:
\eqn\qprime{q_i = V^q_{ij} q^\prime_i}
where $V^q$ satisfy
\eqn\sati{V^q V^{q\dagger} = (R^q)^{-1}.}
(The matrix $R^q$ is hermitian and positive definite.  Therefore, eq.
\sati\ has a solution. The ambiguity in the solution under
multiplication of $V^q$ {}from the right by a unitary transformation can
be fixed by imposing that $V^q$ is hermitian.) The true mass matrices are
then
\eqn\true{M^{\prime d} = (V^Q)^T M^d V^{\bar d} ; \quad
M^{\prime u} = (V^Q)^T M^u V^{\bar u} .}
We should now ask: What are the consequences of the distinctions between
$M$ and $M^\prime$?

To answer the question we have to analyze the situation more carefully.
Clearly,
\eqn\vqeq{V^q= (\sqrt{1 + r^q})^{-1} = 1-{1 \over 2}r^q + \CO((r^q)^2).}
The matrix elements of $V^q$ are of the same order of magnitude as those
of $R^q$.  To show that, we should make sure that no element of a power
of $r^q$ is larger than that of $r^q$.  This fact follows {}from
\eqn\rqij{\eqalign{(r^q)^2_{ij} \sim & \sum_k
\ea^{|H_1(q_i)-H_1(q_k)|}\eb^{|H_2(q_i) - H_2(q_k)|} \ea^{|H_1(q_k)-
H_1(q_j)|}\eb^{|H_2(q_k) - H_2(q_j)|} \cr
& \roughly< \sum_k \ea^{|H_1(q_i) -
H_1(q_j)|}\eb^{|H_2(q_i) - H_2(q_j)|} \sim r^q_{ij}.\cr}}
We conclude that
\eqn\solvr{R^q_{ij} \sim V^q_{ij}.}

Consider now the master model with only a single $U(1)$ and all charges
non-negative.  Without loss of generality we can limit ourselves to
$M^d$:
\eqn\master{M^{\prime d}_{ij}=\sum_{lk} V^Q_{li} M^d_{lk}V^{\bar d}_{kj}
\sim\sum_{lk} \l^{|H(Q_i) - H(Q_l)|} \l^{H(Q_l) + H(\bar d_k)}
\l^{|H(\bar d_j) - H(\bar d_k)|} \sim \l^{H(Q_i) + H(\bar d_j)},}
where in the last step only terms in the sum over $l,k$ with both
$H(Q_i) - H(Q_l) \ge 0$ and $ H(\bar d_j) - H(\bar d_k)\ge 0$
contribute.  We conclude that the numbers of order one in $M$ can
change but the order of magnitude is unchanged.

Next, consider more complicated models with two $U(1)$'s.  By expressing
$\ea$ and $\eb$ in terms of $\l$ and using \master, it is clear that the
order of magnitude of the various entries in $M$ cannot be modified.
The only danger is that we might lift some of the zeros in $M$.  Again,
without loss of generality we can limit ourselves to $M^d$:
\eqn\twouones{
\eqalign{M^{\prime d}_{ij}& = \sum_{lk} V^Q_{li} M^d_{lk}
V^{\bar d}_{kj} \sim
\sum_{lk} \ea^{| H_1(Q_i) - H_1(Q_l)|}\eb^{|H_2(Q_i) - H_2(Q_l)|}\cr
& \ea^{H_1(Q_l) + H_1(\bar d_k)}\eb^{H_2(Q_l) + H_2(\bar d_k)}
\ea^{|H_1(\bar d_j) - H_1(\bar d_k)|} \eb^{|H_2(\bar d_j) - H_2(\bar
d_k)|},\cr} }
where the sum over $l,k$ is restricted to terms with both
\eqn\lkrest{H_1(Q_l) + H_1(\bar d_k) \ge 0; \quad
{\rm and} \quad H_2(Q_l) + H_2(\bar d_k) \ge 0.}
Every term in the sum is smaller than or equal to $\ea^{H_1(Q_i) +
H_1(\bar d_j)}\eb^{H_2(Q_i) + H_2(\bar d_j)} $ with equality only when
\eqn\equacon{\eqalign{
H_1(Q_i) - H_1(Q_l) & \ge 0,\cr
H_2(Q_i) - H_2(Q_l)& \ge 0,\cr
H_1(\bar d_j) - H_1(\bar d_k)& \ge 0,\cr
H_2(\bar d_j) - H_2(\bar d_k)& \ge 0.\cr}}
Using \lkrest\ and \equacon\ it is easy to see that this is possible
only when both $H_1(Q_i) + H_1(\bar d_j) \ge 0$ and $H_2(Q_i) + H_2(\bar
d_j) \ge 0 $, {\it i.e.} only when $M^d_{ij} \not= 0$.  However, in the
case of interest with $M^d_{ij}=0$, all the terms in the sum in
\twouones\ are smaller than the value in the master model $\ea^{H_1(Q_i)
+ H_1(\bar d_j)}\eb^{H_2(Q_i) + H_2(\bar d_j)} $ by at least by one power
of $\max (\ea, \eb)$.

We conclude that the renormalization of the kinetic terms can modify the
numbers of order one but cannot lift the zeros in the mass matrix to
their master model value.

In the aligned model we needed much better accuracy.  The zeros were not
allowed to be lifted to a high power of $\l$.  Fortunately, no more work
is needed in these cases.  The matrices $R^q$ are of the same order as
the squark mass matrices ${1\over\tm^2}\tM^2$,
and the matrices $V^q$ are of the order of the
matrices which diagonalize the squark mass matrices, $\tV^q$.
Therefore, the
true $V_L^q$ and $V_R^q$ which diagonalize the quark mass matrices are
of the same order as the $K$ matrices of equation \mixgluino\ and the
discussion in subsection 2.3 is not modified.

Clearly, when the $U(1)$ symmetries are replaced by discrete symmetries
which lift the zeros a more careful analysis is needed.

Finally, we mention that if the high energy theory is of the FN type,
then further suppression of the deviation of $R^q$ {}from a unit matrix
may occur. Detailed examination of the tree diagrams that lead to
renormalization of the kinetic terms reveals that it is proportional to
at least one power of $\epsilon_i$ (it could be either $\ea$ or $\eb$)
and at least one power of $\epsilon_j^\dagger$ (which, again, could be
either $\epsilon_1^\dagger$ or $\epsilon_2^\dagger$). This suppression
may go well beyond the naive horizontal suppression that we assumed in
this appendix, making the effects of renormalization of kinetic terms
entirely negligible.
\listrefs
\end